\documentclass[a4paper, 12pt]{article}
\usepackage[left=2.5cm,right=2.5cm,top=3cm]{geometry}

\usepackage[doublespacing]{setspace}
\usepackage{amsmath}
\usepackage{amsfonts}
\usepackage{amssymb}
\usepackage{amsthm} 
\usepackage{multicol}
\usepackage{graphicx}
\usepackage{enumerate}
\usepackage[inline,shortlabels]{enumitem}
\usepackage{bbm}
\usepackage[usenames,dvipsnames]{color}
\usepackage[colorlinks=true,linkcolor=blue, citecolor=blue, urlcolor=blue]{hyperref}
\usepackage{fancyhdr}
\usepackage{epsfig}
\usepackage{epstopdf}
\usepackage{caption}
\usepackage[textfont=footnotesize,justification=centering]{subcaption}
\usepackage{multirow}
\usepackage{appendix}
\usepackage{tabularx}
\usepackage{threeparttable}
\usepackage{float}
\usepackage[dvipsnames]{xcolor}

\usepackage{titling}
\usepackage{sectsty}
\usepackage{float}

\newtheorem*{theorem-non}{Theorem}
\newtheorem{theorem}{Theorem}
\newtheorem{definition}{Definition}
\newtheorem{proposition}{Proposition}

\newtheorem{corollary}{Corollary}
\newtheorem{remark}{Remark}

\newtheorem{example}{Example}

\usepackage[round]{natbib}
\def \E{\mathbb{E}}
\def \P{\mathbb{P}}
\def \Q{\mathbb{Q}}
\newcommand{\RNum}[1]{\uppercase\expandafter{\romannumeral #1\relax}}

\DeclareMathOperator*{\esssup}{ess\,sup}
\DeclareMathOperator*{\essinf}{ess\,inf}
\usepackage[utf8]{inputenc}

\providecommand{\keywords}[1]
{
  	
  \textbf{\textit{Keywords ---}} #1
}

\title{Predictable Forward Performance Processes:\\ Infrequent Evaluation and Applications to Human-Machine Interactions}

\author{
Gechun Liang\thanks{Department of Statistics, University of Warwick. Email: \href{mailto:g.liang@warwick.ac.uk}{g.liang@warwick.ac.uk}} 
\and Moris S. Strub\thanks{Department of Information Systems and Management Engineering, Southern University of Science and Technology and Warwick Business School, University of Warwick. Email: \href{mailto:strub@sustech.edu.cn}{strub@sustech.edu.cn}} 
\and Yuwei Wang\thanks{
Corresponding author. Department of Statistics, University of Warwick and Department of Information Systems and Management Engineering, Southern University of Science and Technology. 
Email: \href{mailto:yuwei.wang.2@warwick.ac.uk}{yuwei.wang.2@warwick.ac.uk}}
}

 \date{ }

\begin{document}

\maketitle
\begin{abstract}
We study discrete-time predictable forward processes when trading times do not coincide with performance evaluation times in a binomial tree model for the financial market.
The key step in the construction of these processes is to solve a linear functional equation of higher order associated with the inverse problem driving the evolution of the predictable forward process.  
We provide sufficient conditions for the existence and uniqueness and an explicit construction of the predictable forward process under these conditions. 
Furthermore, we find that these processes are inherently myopic in the sense that optimal strategies do not make use of future model parameters even if these are known.
Finally, we argue that predictable forward preferences are a viable framework to model human-machine interactions occuring in automated trading or robo-advising. 
For both applications, we determine an optimal interaction schedule of a human agent interacting infrequently with a machine that is in charge of trading. 

\end{abstract}  


\section{Introduction}\label{sec:Introduction}
Classical expected utility maximization requires to determine ex ante three basic elements: the investment horizon, the market model, and the performance criterion in terms of a utility function applying at the chosen terminal time. This fundamental setup has, however, two important limitations. First, the investor must pre-specify her future risk preference for evaluating the performance of investment strategies and the market model for describing asset dynamics for the entire investment horizon. 
As a consequence, the risk preference and the market model cannot be adjusted to new market observations over time.
This is problematic, especially when the investment horizon lies in the distant future. Second, the investment horizon needs to be set before the investor enters the market. 

Forward performance processes are an alternative performance criterion that can address these issues. 
Their continuous-time version was introduced in various forms in \cite{musiela2006investments,musiela2008optimal, musiela2009portfolio, musiela2010portfolio}, \cite{henderson2007horizon}, \cite{choulli2007minimal}, \cite{vzitkovic2009dual} and \cite{mohamed2013exact},
and further developed in, for example,
\cite{Avanesyan2020},
\cite{chong2019pricing}, 
\cite{Bo2022}, \cite{el2014, el2018consistent,el2021},  \cite{el2021recover},\cite{he2021forward}, \cite{hu2020systems}, \cite{kallblad2018dynamically}, \cite{k2020black},
\cite{liang2017representation}, \cite{nadtochiy2017optimal}, and \cite{shkolnikov2016asymptotic}.

In contrast, the discrete-time case is less well understood. 
To the best of our knowledge,  the only studies concerned with the analysis thereof are  \cite{angoshtari2020predictable}, where the framework was first introduced, \cite{strub2021evolution} who extend some of the key results therein to more general models for the financial market and investigate the associated dynamics of risk preferences, and the recent \cite{angoshtari2022predictable}, who establishes existence results in general complete markets and a new solution method for the generalized integral equations associated with the construction of discrete-time, predictable forward processes based on the Fourier
transform for tempered distributions.

An advantage of the discrete-time formulation of forward performance processes is that they are \textit{predictable} instead of just adapted. 
This leads to a more intuitive relation of the utility functions at two consecutive time points. 
We herein build on the work of \cite{angoshtari2020predictable},  and aim to extend their key results to the multi-period binomial tree model for the financial market. 
A key feature, both conceptually and technically, of this extension is that performance evaluation times generally do not coincide with trading times, but occur at a lower frequency. 
This setting is of particular relevance for wealth management, where interaction with the client often occurs at a lower frequency than trading. 

According to the general scheme developed in \cite{angoshtari2020predictable}, the key step in the construction of a predictable forward process is to solve an associated \textit{inverse} investment problem, where one is given an initial utility function and model for the market and seeks to determine a utility function applying at terminal time such that the initial utility function becomes the value function of the resulting expected utility maximization problem.  
Whereas this is a single-period problem in the binomial case studied in \cite{angoshtari2020predictable}, we herein face a multi-period inverse investment problem.
Because the financial market for each single evaluation period is complete, the results of \cite{strub2021evolution} apply and a solution to the multi-period inverse investment problem can be obtained by solving an associated generalized integral equation. 
In the binomial tree model considered herein, the associated generalized integral equation is a linear functional equation of higher order. 
Our main technical contributions are sufficient conditions for existence and uniqueness for the associated equation as well as an explicit construction of a solution under those conditions.   
An overview of the general theory of functional equations can be found for example in \cite{kuczma_choczewski_ger_1990}, \cite{kress1989linear}, \cite{polyanin2008handbook}, or \cite{zemyan2012classical}. 
There are interesting applications of this theory in fields as diverse as geometry, probability theory, financial management, or information theory.

An interesting observation is that optimal strategies associated with a predictable discrete-time forward performance process are inherently \textit{myopic} in the sense that they do not make use of information about future market parameters.
This is in stark contrast to the classical, backward expected utility maximization problem where optimal strategies generally depend on future market parameters or characteristics thereof. 
Another observation is that discrete-time predictable forward performance processes are decreasing in the evaluation period. 
In continuous time, forward performance processes are not necessarily monotone in time. 
However, continuous-time forward performance processes that are time-monotone often allow for more explicit results, see, e.g., \cite{musiela2009portfolio} and \cite{rogers2009characterization}.

The second major contribution of this paper is an application of  discrete-time predictable forward processes with infrequent evaluation as a framework to model human-machine interactions such as automated trading and robo-advising.
To the best of our knowledge, this is the first application of the theory of forward preferences to the asset allocation problems faced by an automated trading system or a robo-advisor.

Automated trading dates back as far as the 1970s and was developed out of the introduction of designated orders turnaround system, see \cite{grossman1988program}.
Broadly defined, automated trading refers to the execution of orders by an algorithm according to a pre-defined trading strategy. 
To date, automated trading systems are widely used by institutional and retail traders alike.
The literature on automated trading strategies is vast, see for example the monographs  \cite{cartea2015algorithmic} and \cite{aldridge2013high}. 
Another ongoing and controversial topic of research is studying the impact of automated trading on financial markets.
For example,  \cite{hendershott2011does} found that automated trading can improve market efficiency and liquidity, \cite{chaboud2014rise} and \cite{brogaard2014high} argue that automated trading improves price and informational efficiency, and the implications on behaviors and strategies of traders are investigated in \cite{o2015high}.

Robo-advisors constitute a class of wealth management tools that offer asset allocation recommendations and implementations based on algorithms and automated by software, see, for example, \cite{DAcuntoRossi21:Inbook} for an overview and taxonomy. 
They contribute to the democratization of finance by making wealth management services that were previously limited to a select group of wealthy investors available for all. 
Since emerging in the late 2000's, robo-advising services have experienced rapid growth and are now estimated to manage over USD $1,600$ billion of over $500$ million clients globally. 

In both applications, we consider a machine trading on behalf of a human agent at a high frequency and interacting with the agent at a lower frequency. 
In the case of automated trading, the human agent is a market expert that periodically communicates updated assessments of the market as inputs for an automated trading system (ATS). 
In the case of robo-advising, the human agent is a client that periodically communicates her risk-preferences to the robo-advisor.  
Predictable forward processes with infrequent evaluation have three important features making them expedient for such applications.

First, the construction of forward processes assures that optimal investment strategies are time-consistent. 
This is in stark contrast to the dynamic mean-variance objective.  
Whenever preferences are time-inconsistent, one has to decide on whether to work with pre-committed or equilibrium strategies, and there does not seem to be a canonic choice for the two applications. 
For example, when modeling the asset allocation problem of a robo-advisor,
\cite{capponi2022personalized} and \cite{dai2021dynamic} work with equilibrium strategies while \cite{cui2022risk} introduce  mean-variance induced utility functions to avoid the issue altogether. 
However, it seems also plausible to work with pre-committed strategies and regard the machine (namely, the automated trading system or the robo-advisor) as a pre-committment device. 
Working with forward processes avoids this discussion and leads to strategies that are globally optimal. 

Second, forward processes accommodate dynamically changing investment horizons. 
While this feature is an advantage for portfolio selection in general, it is of particular relevance for our applications. 
Imagine a situation where the investment horizon the human agent is reached, but the agent forgets to withdraw her funds or otherwise communicate with the trading platform. 
How should the trading platform act in this situation if it aims to continue investing in the best interest of the human agent? 
Forward preferences provide an elegant solution to this problem: Continue investing in a manner that is consistent with previous preferences and decisions by updating preferences according to the martingale optimality principle.  


Third, in addition to these general advantages of forward performance processes, the specific class we investigate herein allow for the additional feature that trading times do not necessarily coincide with performance evaluation times. 
This is of practical relevance for the applications we have in mind, as trading typically occurs at a a higher frequency than interaction with the human agent. 

In the automated trading application, we consider a human expert operating an ATS whose preferences are described by a discrete-time predictable performance process. 
The expert gathers information about the time-varying parameters describing the financial market and updates these at infrequently occurring interactions with the ATS. 
On the other hand, the ATS manages the portfolio on behalf of the expert period-by-period based on the assessment of the market communicated by the expert at the last interaction time.
The expert seeks to determine an optimal schedule for interacting with the ATS that balances a tradeoff between accuracy about the current values of the market parameters and a cost incurred when assessing the market. 
We characterize the optimal interaction schedule and find that it balances a tradeoff between the cost required to assess the market parameters  and expected loss in performance due to the inaccuracy about the market parameters. 
As one could intuitively expect, the optimal interaction schedule is increasing in the interaction cost and decreasing in a uniform increase of uncertainty about the market parameters.
However, the effect of a non-uniform increase in the uncertainty is more intricate, and it can indeed happen that the optimal interaction schedule increases when uncertainty about the market parameters in the near future increases. 
This occurs because an increase in the uncertainty about market parameters in the near future harms performance after each interaction time. 
Interacting more frequently therefore does not necessarily lead to better performance. 
We also numerically investigate how the optimal interaction schedule depends on the risk-aversion of the human expert. 
Typically, a more risk-averse expert is interacting more frequently with the ATS than a less risk-averse expert. 
However, when the interaction cost is large and either the expected return of the risky asset is close to the risk-free return or the risk-aversion is already large, then an increase in risk-aversion can lead to an increase of the optimal interaction schedule. 
In this case, the investment in the risky asset is very small, and the updating of the probability for a positive outcome does not lead to a significant change in optimal investment strategies. 

In the robo-advising application, we consider a client of a robo-advisor whose preferences are described by a general stochastic utility process.
The robo-advisor manages the portfolio on behalf of the client, but only has accurate knowledge about the client's risk preferences whenever there is an interaction. 
The robo-advisor also has an accurate understanding of the parameters specifying the current financial market. 
But as associated optimal strategies have to be approved by the client, market parameters are effectively only updated whenever there is an interaction as well. 
We seek to determine an optimal schedule of interaction between robo-advisor and client that bounds the deviation between the actual strategy implemented by the robo-advisor and an ideal strategy that would be obtained if the robo-advisor always has accurate knowledge about the risk preferences of the client and can update market parameters period-by-period.
We characterize an optimal interaction schedule under a robust approach and find that the client should interact more frequently when she is less tolerant about deviations from an ideal strategy or when there is greater uncertainty about market or preference parameters. 
Interestingly, we also show that it is optimal to interact less frequently in a more volatile market environment all else being equal. 
The intuition is that both the ideal strategy and the strategy implemented by the robo-advisor are less aggressive in a volatile environment and their deviation consequentially smaller. 
A further numerical analysis indicates that the optimal interaction schedule is more sensitive about the uncertainty in market parameters than about the uncertainty in preference parameters.

The remainder of this paper is organized as follows.
In Section \ref{sec:Model}, we introduce the model for the financial market and review the definition and preliminary results for discrete-time predictable performance processes.
We provide sufficient conditions for existence and uniqueness and an explicit construction of the discrete-time predictable forward process in Section \ref{sec:FunctionalEquation}.
In Sections \ref{sec:AutomatedTrading} and \ref{sec:RoboAdvising}, we discuss discrete-time predictable forward processes as a potential framework to model preferences for automated trading and robo-advising applications. 
Section \ref{sec:Conclusions} concludes the paper. 

 \section{Discrete-time predictable forward performance processes: Model and definition}\label{sec:Model}


In this section, we introduce the notion of discrete-time predictable forward performance process with evaluation period larger than one in a binomial tree model which was originally presented in \cite{cox1979option} for option pricing. Discrete-time predictable forward performance processes were introduced in \cite{angoshtari2020predictable} for general models of the financial market. 
However, their analysis is limited to the single-period binomial model where trading dates and performance evaluation dates coincide.  
The complete semimartingale model in \cite{strub2021evolution} is more general than the setup of this paper, but they do not provide conditions for existence and do not explicitly construct discrete-time predictable forward processes as we will herein. 

The investor starts at time zero with preferences over wealth represented by a utility function $U_0$.
We herein assume that any utility function $U: \mathbb{R}^+ \rightarrow \mathbb{R}$ is twice continuously differentiable, strictly increasing, strictly concave and satisfies the Inada conditions.
We fix a probability space $(\Omega, \mathcal{F}, \P)$, where $\P$ denotes the real (historical) probability measure on ($\Omega, \mathcal{F}$). 
Throughout the paper, $\mathbb{N}$ denotes the set of positive integers and $\mathbb{N}_0$ is the set of nonnegative integers.

We suppose that the investor trades at discrete times $n$, $n \in \mathbb{N}_0$, between a risk-free bond whose (discounted) price offers zero interest and a single risky stock. 
The (discounted) price process $S = (S_n)_{n \in \mathbb{N}_0}$ of the stock is described by a binomial model
\begin{align*}
    S_n = S_{n-1}\left( u_n B_n + d_n (1-B_n) \right), \quad n \in \mathbb{N},
\end{align*}
and $S_0 = 1$, where $B_n \in \{0,1\}$ for all $n \in \mathbb{N}$, i.e., $(B_n)_{n \in \mathbb{N}}$ is a sequence of Bernoulli random variables.  
We allow for the market parameters $(u_n)_{n \in \mathbb{N}}$, $(d_n)_{n \in \mathbb{N}}$, and $(p_n)_{n \in \mathbb{N}}$ to be stochastic processes satisfying $d_n, p_n \in (0,1)$ and $u_n > 1$, $n \in \mathbb{N}$. 
For methods to calibrate binomial models we refer to \cite{cox1979option}, \cite{Rubinstein94:JF}, or \cite{Jackwerth99:JoD}. 
Here and throughout the paper, we assume that all relations hold $\mathbb{P}$-almost surely. 
The investor evaluates her portfolio at performance evaluation times $(\tau_k)_{k \in \mathbb{N}_0}$ given by $\tau_k = km$, where $m \in \mathbb{N}$ is the \textit{evaluation period length}.
One could more generally consider $(\tau_k)_{k \in \mathbb{N}_0}$ to be a stochastic process
 taking values in $\mathbb{N}_0$ such that $\tau_0=0$ and $\tau_{k+1} > \tau_k$
and a measurability requirement implying that the length of each evaluation period is known at the beginning of the respective period. 

We specify the filtration $\mathbb{F} = (\mathcal{F}_n)_{n \in \mathbb{N}_0}$ by supposing that $\mathcal{F}_n$ is the augmented $\sigma$-algebra generated by $(B_j)_{j=1}^n$ and $(u_j,d_j,p_j)_{j=1}^{km}$ with $k$ such that $(k-1)m \leq n \leq km - 1$. This specification of the filtration is to be interpreted as follows. 
At any point in time, the investor knows the past price levels of the stock and the market parameters of the past and current (performance) evaluation period. 
However, the market parameters of subsequent evaluation periods remain stochastic. 
We complete the specification of the market by assuming that 
\begin{align*}
    \mathbb{P} \left[ B_n = 1 \vert \mathcal{F}_n \right] = 1 -  \mathbb{P} \left[ B_n = 0 \vert \mathcal{F}_n \right] = p_n, \qquad n \in \mathbb{N}.
\end{align*}
This assures that $p_n$ satisfies the usual interpretation of the conditional probability of an upward move of the stock in the $n$'th trading period.

Performance evaluation occurring less frequently than trading and the enlargement in filtration are the exact differences between the model studied herein and \cite{angoshtari2020predictable}.
When $m=1$, trading times and performance evaluation times coincide and the model reduces to the one extensively studied therein. 
However, in general, the evaluation period length is strictly larger than one and trading thus occurs at a higher frequency than performance evaluation. 
This separation between trading times and performance evaluation times is a key feature of our model and will be at the heart of our analysis and applications. 
We remark that we make an implicit assumption that trading is more frequent than performance evaluation and that the investor can trade at every performance evaluation time. 
This is natural.
Performance evaluation without concurrent trading would not be observable. 

Trading strategies are described by means of predictable processes
$\pi = \left( \pi_n \right)_{n \in \mathbb{N}}$, where $\pi_n$ denotes the dollar amount invested in the risky asset over trading period $[n-1, n)$. 
A portfolio is constructed by following the trading strategy on the stock while investing all the remaining wealth in the risk-free bond. 
Given an initial wealth $x >0$ and self-financing trading strategy $\pi$, the wealth process $X^\pi = \left(X_n^{\pi}  \right)_{n \in \mathbb{N}}$ evolves according to
$X_n^{\pi} = x + \sum\nolimits_{i=1}^n\pi_i\left(\frac{S_i}{S_{i-1}} -1 \right)
$. A trading strategy $\pi$ as well as the associated wealth process $X^{\pi}$ are called \textit{admissible} if $X^{\pi}$ is nonnegative. 
We denote by $\mathcal{A}(n,x)$ the set of admissible trading strategies $\left( \pi_k \right)_{k \geqslant n}$ and by $\mathcal{X}(n,x)$ the associated wealth processes $\left( X^{\pi}_k \right)_{k \geqslant n}$ starting from $X^{\pi}_n = x$, $n \in \mathbb{N}$, and abbreviate $\mathcal{A}(0,x)$, $\mathcal{X}(0,x)$ by $\mathcal{A}(x)$, $\mathcal{X}(x)$. We often drop the explicit dependence of a wealth process on the trading strategy and write $X \in \mathcal{X} (n,x)$. 

We remark that the model for the financial market described above is \textit{sequentially complete across each evaluation period} in the following sense. 
For any $k \in \mathbb{N}$ and $\mathcal{F}_{km}$-measurable random variable $X_m \geq 0$ there exist an $\mathcal{F}_{(k-1)m}$-measurable random variable $X_{(k-1)m}$ such that $X_m \in \mathcal{X} (m, X_{(k-1)m})$. 
In other words, any random variable measurable with respect to the filtration at the end of an evaluation period can be generated by admissible trading starting from a random variable measurable at the beginning of the same evaluation period. 
This feature is a straightforward consequence from the fact that model parameters are known at the beginning of each evaluation period and that, thus, each evaluation period in isolation is nothing but a standard binomial model. 
However, the model is not necessarily complete across multiple evaluation periods. 
For example, an $\mathcal{F}_{2m}$-measurable random variable $X_{2m}$ is not necessarily replicable by admissible trading from a deterministic initial wealth. 
This is because the model parameters over the second evaluation period $(m, 2m]$ are not known at the beginning of the first evaluation period. 
Hence, $X_{2m}$ cannot be hedged by admissible trading if the model parameters are determined in part through some exogenous random noise.
The market would become complete if market parameters are adapted to the filtration generated by the stock price. 
We study an example of such a market in greater detail in Subsection \ref{MLEExample} where we consider a specific rule of how market parameters are updated in response to previous outcomes of the stock.

We next present the definition of discrete-time predictable forward performance processes with evaluation period length $m$.

\begin{definition}\label{def:PredictableForwardProcess}
A family of random functions $\left\lbrace U_{km} : \mathbb{R}^+ \times \Omega \rightarrow \mathbb{R} \vert k \in \mathbb{N}_0 \right\rbrace$ is called a discrete-time predictable forward performance process with evaluation period length $m \in \mathbb{N}$ (an $m$-forward process in short) if the following conditions hold:
\begin{enumerate}
\item[(i)] $U_0(x,\cdot)$ is constant and $U_{km}(x,\cdot)$ is $\mathcal{F}_{(k-1)m}$-measurable for each $x\in\mathbb{R}^+$ and $k \in \mathbb{N}$. 
\item[(ii)] $U_{km} (\cdot ,\omega )$ is a utility function for almost all $\omega \in \Omega$ and all $k \in \mathbb{N}_0$.
\item[(iii)] For any initial wealth $x > 0$ and admissible wealth process $X \in \mathcal{X}(x)$,
\begin{align*}
U_{(k-1)m} \left( X_{(k-1)m} \right) \geqslant \E \left[ U_{km} \left( X_{km} \right) \big\vert \mathcal{F}_{(k-1)m} \right], \quad k \in \mathbb{N}.
\end{align*}
\item[(iv)] For any initial wealth $x>0$, there exists an admissible wealth process $X^* \in \mathcal{X} (x)$ such that 
\begin{align*}
U_{(k-1)m} \left( X^*_{(k-1)m} \right) = \E \left[ U_{km} \left( X^*_{km} \right) \big\vert \mathcal{F}_{(k-1)m} \right], \quad k \in \mathbb{N}.
\end{align*}
\end{enumerate}
\end{definition}

Definition \ref{def:PredictableForwardProcess} is analogous to its single-period counterpart, but we are now interested in the case where trading occurs more often than performance evaluation. See \cite{angoshtari2020predictable} for a detailed discussion of the definition and a theoretical framework of discrete-time predictable forward performance processes. 
Considering discrete-time predictable forward performance process with evaluation period larger than one is more general mathematically and also relevant for applications. 
It is increasingly the case that trading is automated and executed by machines at a higher frequency than monitoring and analyzing of the investment portfolio by a human agent. 
Modelling a framework where trading occurs at a higher frequency than performance evaluation and preference updating is thus important for investment practice. 

Property $(i)$ requires that preferences applying at the end of an evaluation period are known at the beginning of that period. This reflects the \textit{predictability} of discrete-time predictable forward processes adapted to multi-period evaluation of the performance. 
Properties $(iii)$ and $(iv)$ demand that an $m$-forward process evolves under the guidance of Martingale Optimality Principle and ensure time-consistency of optimal strategies. 
In addition, properties $(iii)$ and $(iv)$ imply that
\begin{align}\label{eq:InverseInvestment_Dynamic}
U_{(k-1)m} \left( X^*_{(k-1)m} \right) = \esssup_{X_{km} \in \mathcal{X} \left((k-1)m,X^{*}_{(k-1)m}\right)} \E \left[ U_{km}  \left(X_{km}\right) \bigg\vert \mathcal{F}_{(k-1)m}\right].
\end{align}
Iteratively solving \eqref{eq:InverseInvestment_Dynamic} leads to the construction of the $m$-forward process, see \cite{angoshtari2020predictable} for a detailed exposition. 
The crucial step is to solve the following \textit{inverse investment problem}:
Given an initial utility function $U_0$, we seek for a forward utility function $U_m$ such that for any $x > 0$,
\begin{align}\label{eq:InverseInvestmentProblem}
U_0(x) = \sup_{X_{m} \in \mathcal{X}(x)}\E \left[ U_m \left( X_m \right)\right]=\sup_{\pi \in \mathcal{A}(x)}\E \left[ U_m \left( x + \sum\nolimits_{i=1}^m\pi_i(R_i-1) \right)\right]. 
\end{align}
One can then construct $U_{2m}, U_{3m},...$ by repeatedly solving a problem of the form \eqref{eq:InverseInvestmentProblem} conditionally on updated information available at next evaluation point and arguing that this solution satisfies the required measurability conditions. 
We emphasize that obtaining a solution that is measurable as a function of the market parameters is necessary for the construction of a predictable forward process, cf. \citet[Remark 2.2]{strub2021evolution} for details.

\begin{remark}
When deriving the solution to the inverse investment problem \eqref{eq:InverseInvestmentProblem}, we will carefully argue that the constructed forward utility function depends in a measurable way on all market parameters at the previous evaluation time, and that this will allow us to obtain a predictable process.
Therefore, the dynamic version of the sequence of random problems \eqref{eq:InverseInvestment_Dynamic} can be reduced to the deterministic version \eqref{eq:InverseInvestmentProblem}. 
\end{remark}

In analogy to the terminology in \cite{strub2021evolution}, we will refer to an initial utility function $U_0$ and a utility function $U_m$ solving \eqref{eq:InverseInvestmentProblem} as an \textit{m-forward pair} $(U_0,U_m)$.
Note that our assumptions imply that the model input is known at the beginning for the evaluation period as a deterministic triplet $\left( (p_i)_{i=1,\dots, m}, (u_i)_{i=1,\dots, m}, (d_i)_{i=1,\dots, m} \right)$.
Recall from the above discussion that the market is sequentially complete across each evaluation period and that the equivalent martingale measure for the truncated model of a single evaluation period is therefore unique. 
We denote it by $\mathbb{Q}$ and let $q_i = \frac{1-d_i}{u_i - d_i}$, $i = 1, \dots, m$, be the risk-neutral probability for an upward move of the stock in the $i$'th trading period.

A key result for the theory of discrete-time predictable forward processes is the equivalence between the inverse investment problem \eqref{eq:InverseInvestmentProblem} and a generalized integral equation for the inverse marginal or the conjugate corresponding to the involved forward pair. This was shown for the binomial market in \cite{angoshtari2020predictable} and generalized to complete semimartingale models in \cite{strub2021evolution}. 
To state this result, we recall the definition of an inverse marginal function.
 An \textit{inverse marginal function} $I: \mathbb{R}^+ \rightarrow \mathbb{R}^+$ is continuously differentiable, strictly decreasing and satisfies $\lim_{y \to +\infty} I(y)=0$ and $\lim_{y \to 0^{+}} I(y)=\infty$. For a given utility function $U(x), x \in \mathbb{R}^+$, $I(y)=(U^{'})^{-1}(y)$ is the inverse marginal function corresponding to $U(x)$. We denote the set of utility functions by $\mathcal{U}$, the set of inverse marginal functions by $\mathcal{I}$.
According to Therem 2.4 in \cite{strub2021evolution}, see also Theorems 5.1 and 5.2 in \cite{angoshtari2020predictable} for an earlier version in the single-period binomial setting, solving the inverse investment problem \eqref{eq:InverseInvestmentProblem} in the space $\mathcal{U}$ of utility functions is equivalent to finding a solution to
 \begin{align}\label{eq:FunctionalEquation}
 I_0 (\hat{y}) = \mathbb{E}_{\mathbb{Q}} \left[ I_m \left( \hat{y} \frac{d\mathbb{Q}}{d \mathbb{P}}\right) \right], \quad \hat{y}>0, 
 \end{align}
 in the space $\mathcal{I}$ of inverse marginal functions in the following sense: If $(U_0,U_m)$ is an $m$-forward pair solving \eqref{eq:InverseInvestmentProblem}, then the associated inverse marginal functions $(I_0,I_m)$ solve \eqref{eq:FunctionalEquation}. Vice versa, if $(I_0,I_m)$ is a pair of inverse marginal functions satisfying \eqref{eq:FunctionalEquation}, then the associated utility functions satisfy \eqref{eq:InverseInvestmentProblem} up to a constant, i.e., there is a constant $c \in \mathbb{R}$, which can be expressed explicitly in terms of $U_0$, $I_m$, and the market parameters,  such that $\tilde{U}_m (x) := U_m (x) + c$ satisfies \eqref{eq:InverseInvestmentProblem}. 
Because it is often the case that finding a solution to the generalized integral equation \eqref{eq:FunctionalEquation} is considerably easier than solving the inverse investment problem \eqref{eq:InverseInvestmentProblem}, the generalized integral equation \eqref{eq:FunctionalEquation} plays an important role in the theory of discrete-time predictable forward processes. 
Our main technical contribution is to provide a solution to \eqref{eq:FunctionalEquation} for the binomial market when trading times do not coincide with performance evaluation times, and thus \eqref{eq:FunctionalEquation} reduces to a linear functional equation as in \cite{angoshtari2020predictable} but of higher order. 
Solving \eqref{eq:FunctionalEquation}, together with a thorough analysis of the result, will be the content of the following Section \ref{sec:FunctionalEquation} for the case of time-homogeneous and time-heterogeneous market parameters respectively.

\section{The linear functional equation of higher order}\label{sec:FunctionalEquation}

In this section, we first develop a general approach to solving the linear functional equation \eqref{eq:FunctionalEquation} associated with the inverse investment problem \eqref{eq:InverseInvestmentProblem}. 
We then study a special market setting with \textit{time-homogeneous market parameters}.
This slight loss of generality will allow us to derive more explicit and interpretable results.

\subsection{The heterogeneous case}
In the general case, the agent has possibly heterogeneous beliefs on future price movements across the trading periods constituting a given evaluation period. 
Given the deterministic triplet $\left( (p_i)_{i=1,\dots, m}, (u_i)_{i=1,\dots, m}, (d_i)_{i=1,\dots, m} \right)$ characterizing the multi-period binomial tree,  we follow \cite{angoshtari2020predictable} and set $a_i=\frac{1-p_i}{p_i}\frac{q_i}{1-q_i}$, $b_i=\frac{1-q_i}{q_i}$, $c_i=\frac{1-p_i}{1-q_i}$ for $i=1,2,...,m$, where $q_i = \frac{1-d_i}{u_i - d_i}$ is the risk-neutral probability for an upward move of the stock in the $i$'th trading period. 
Observe that there are $2^m$ possible outcomes for the $m$-period binomial tree with heterogeneous market parameters.
When ordered from the lowest price level to the highest, they occur with probabilities $\prod \limits_{i=1}^mp_i^{\gamma_{j,i}}(1-p_i)^{1-{\gamma_{j,i}}}$, $j = 0,1,...,2^m-1$, where $\gamma_{j,i}$ is defined as the $i'th$ digit of the binary representation of $j$, i.e., $(j)_{10}= (\gamma_{j,m}...\gamma_{j,2} \gamma_{j,1})_2$ and zeros are filled in the front of the binary representation if it contains less than $m$ digits. 
In the current setting, the generalized integral equation \eqref{eq:FunctionalEquation} can thus be written as the linear functional equation
\begin{align}\label{eq:FctEQ_Multinomial_Heterogeneous}
I_0 (\hat{y}) = \sum\nolimits_{j=0}^{2^m-1}\prod \limits_{i=1}^mq_i^{\gamma_{j,i}}(1-q_i)^{1-{\gamma_{j,i}}}I_m \left(\frac{\prod \limits_{i=1}^mq_i^{\gamma_{j,i}}(1-q_i)^{1-{\gamma_{j,i}}}}{\prod \limits_{i=1}^mp_i^{\gamma_{j,i}}(1-p_i)^{1-{\gamma_{j,i}}}} \hat{y} \right) .
\end{align}
Analyzing \eqref{eq:FctEQ_Multinomial_Heterogeneous} is challenging because
the argument of $I_m$ can in general not be transformed to an iterative form. 
However, we are still able to characterize solutions to \eqref{eq:FctEQ_Multinomial_Heterogeneous} within the class of inverse marginal functions and provide conditions for the uniqueness of such a solution. 
This will be the main technical contribution of this section. 

\medskip

For a given initial utility function $U_0 \in \mathcal{U}$ and associated inverse marginal function $I_0 \in \mathcal{I}$ we define the following auxiliary functions,
\begin{align}\label{eq:DefPhiPsiZero_heter}
\Phi_0(y) = I_0(a_1c_1y) - b_1I_0(c_1y)  \quad \mathrm{and} \quad \Psi_0(y) = y^{-\rm{log}_{a_1}b_1}I_0(c_1y), \quad y > 0 ,
\end{align}
and
\begin{align*}
\begin{split}
\Phi_i^{(v_1,\dots,v_i)}(y) & =\frac{\prod \limits_{l=1}^i(1+b_l)}{\prod \limits_{j=1}^ib_j^{v_j}}\bigg(\sum_{n_1=0,...,n_i=0}^{\infty}(-1)^{p_{(n_1,\dots,n_i)}}Q_{(v_1,\dots,v_i);(n_1,\dots,n_i)}I_0\left(R_{(v_1,\dots,v_i);(n_1,\dots,n_i)}a_{i+1}y\right)\\ & \hspace{2.2cm} -b\sum_{n_1=0,...,n_i=0}^{\infty}(-1)^{p_{(n_1,\dots,n_i)}}Q_{(v_1,\dots,v_i);(n_1,\dots,n_i)}I_0\left(R_{(v_1,\dots,v_i);(n_1,\dots,n_i)}y\right) \bigg),
\end{split}
\end{align*}
\begin{align}\label{eq:DefPhiPsi_heter}
\begin{split}
\Psi_i^{(v_1,\dots,v_i)}(y) & = y^{-({\rm{log}}_{a_{i+1}}{b_{i+1}})}\frac{\prod \limits_{l=1}^i(1+b_l)}{\prod \limits_{j=1}^ib_j^{v_j}}\\
& \qquad \times \sum\limits_{n_1=0,...,n_i=0}^{\infty}(-1)^{p_{(n_1,\dots,n_i)}}Q_{(v_1,\dots,v_i);(n_1,\dots,n_i)}I_0\left(R_{(v_1,\dots,v_i);(n_1,\dots,n_i)}y\right),
\end{split}
\end{align}
for $y>0$, $i=1,\dots,m-1$, and $(v_1,\dots,v_i) \in \{0,1 \}^{i}$, where
  $Q_{(v_1,\dots,v_i);(n_1,\dots,n_i)}=\prod\limits_{k=1}^ib_k^{n_k(1-2v_k)}$, $R_{(v_1,\dots,v_i);(n_1,\dots,n_i)}=\prod\limits_{s=1}^ia_s^{n_s(2v_s-1)+(v_s-1)}\prod\limits_{u=1}^{i+1}c_u$ and $ p_{(n_1,\dots,n_i)}=\sum\nolimits_{k=1}^{i}n_k$. For a given pair of functions $(\Phi,\Psi)$, we say that the pair satisfies condition $(C1)$ if 
\begin{align*}
    \Phi^{'}(y) > 0 \ {\rm{and\ either}}\ a>1\ {\rm{and}}\ \lim_{y\to\infty} \Psi (y) = 0 \ {\rm{or}}\ a<1 \ {\rm{and}}\  \lim_{y\to0^+} \Psi (y) = 0.
\end{align*}
We say that the pair of functions $(\Phi,\Psi)$ satisfies condition $(C2)$ if
\begin{align*}
    \Phi^{'}(y) < 0 \ {\rm{and\ either}}\ a>1\ {\rm{and}}\ \lim_{y\to0^+} \Psi (y) = 0 \ {\rm{or}}\ a<1 \ {\rm{and}}\  \lim_{y\to\infty} \Psi (y) = 0.
\end{align*}
Next, we iteratively define a sequence $\{ (\alpha_1 , \dots , \alpha_i )\}_{i = 1 ,\dots, m}$ starting with $(\alpha_1)=(1)$ if $\left( \Phi_{0}, \Psi_{0} \right)$ satisfies $(C1)$ or $(\alpha_1)=(0)$ if $\left( \Phi_{0}, \Psi_{0} \right)$ satisfies $(C2)$. 
We then define 
\begin{align*}
    (\alpha_1,\dots,\alpha_{i+1}) = \left\{ \begin{array}{ll}
        (\alpha_1,\dots,\alpha_i,1) & \mathrm{if} \enskip \left( \Phi_{i}^{(\alpha_1,\dots,\alpha_i)}, \Psi_{i}^{(\alpha_1,\dots,\alpha_i)} \right) \enskip \mathrm{satisfies} \enskip (C1), \\
     (\alpha_1,\dots,\alpha_i,0)    & \mathrm{if} \enskip \left( \Phi_{i}^{(\alpha_1,\dots,\alpha_i)}, \Psi_{i}^{(\alpha_1,\dots,\alpha_i)} \right) \enskip \mathrm{satisfies} \enskip (C2),
    \end{array} \right.
\end{align*}
for $i= 1, \dots m-1$. 

\medskip

The following theorem constitutes the main result of this section and  provides an explicit expression for $I_m$ in terms of $I_0$ with their corresponding utility functions being an $m$-forward pair. 
This expression in turn leads to a \textit{construction method} for the $m$-forward pair. 
This theorem is the multi-period analogue to the single-period result in \cite[Theorem 7.1]{angoshtari2020predictable}.

\begin{theorem}\label{thm:mForwardConstructionHeterogeneous}
 Let $U_0 \in \mathcal{U}$ be a utility function with associated inverse marginal function $I_0$ and suppose that $\{(\alpha_1,\dots,\alpha_i)\}_{i=1,\dots,m}$ exists. Define $I_m$  by
\begin{align}\label{eq:ForwardInverseMarginal_Heterogeneous}
    I_m(y)=\frac{\prod \limits_{i=1}^m(1+b_i)}{\prod \limits_{j=1}^mb_j^{\alpha_j}}\sum\limits_{n_1=0,...,n_m=0}^{\infty}(-1)^{p_{(n_1,\dots,n_m)}}\prod\limits_{k=1}^mb_k^{n_k(1-2\alpha_k)}I_0 \left(\prod\limits_{s=1}^ma_s^{n_s(2\alpha_s-1)+(\alpha_s-1)}\prod\limits_{u=1}^mc_uy \right), 
\end{align} 
and
\begin{align*}
    U_m(x):=U_0(1)+ \mathbb{E}_{\mathbb{P}}\left[ \int_{I_m(\frac{d\mathbb{Q}}{d\mathbb{P}} U_0'(1))}^{x} I_m^{-1}(t)dt\right], \quad x>0.
\end{align*}
Then $U_m$ is the unique utility function solving \eqref{eq:InverseInvestmentProblem} and $I_m$ is the unique inverse marginal function solving the generalized integral equation \eqref{eq:FctEQ_Multinomial_Heterogeneous}. 
Moreover, the optimal wealth solving \eqref{eq:InverseInvestmentProblem} is given by 
\begin{align}\label{OptimalWealthHetero}
X_m^*(x)=I_m \left( U_0'(x) \frac{d\mathbb{Q}}{d\mathbb{P}} \right)
\end{align}
%
%
\end{theorem}

We remark that the introduction of these auxiliary functions is to help to establish the uniqueness and existence conditions of the solutions in the class of inverse marginal functions, cf. the proof of Theorem \ref{thm:mForwardConstructionHeterogeneous} for details.
Whether $\{(\alpha_1,\dots,\alpha_i)\}_{i=1,\dots,m}$ exists needs to be determined on an ad hoc basis given an initial datum. 
For example, when the initial datum belongs to a family of CRRA utility functions, $U_0(x) = \log x$, $x>0$ and $U_0(x)=(1-\frac{1}{\theta})^{-1}x^{1-\frac{1}{\theta}}$, $x>0$, where $1\neq \theta>0$, then $\{(\alpha_1,\dots,\alpha_i)\}_{i=1,\dots,m}$ is typically well defined. 
Exceptions are the cases where $p_i=\frac{1}{2}$ or $\theta =- \log_{a_i}{b_i}$.
In these cases, $\left( \Phi_{i}^{(\alpha_1,\dots,\alpha_i)}, \Psi_{i}^{(\alpha_1,\dots,\alpha_i)} \right)$ satisfy neither $(C1)$ nor $(C2)$, but one can still provide a natural candidate for the forward process within the family of power and log utilities and show that this is indeed a forward process. However, uniqueness generally does not hold in this case, cf. \cite[Example 6.1]{angoshtari2020predictable}.
Therefore, we emphasize that the condition that $\{(\alpha_1,\dots,\alpha_i)\}_{i=1,\dots,m}$ exists is sufficient, but not necessary for the existence and uniqueness of the forward process. 
Another example will be treated in Example \ref{ex:SumOf2Power_Homogeneous}. 
How to solve the corresponding functional equation and construct the forward performance process when $\{(\alpha_1,\dots,\alpha_i)\}_{i=1,\dots,m}$ does not exist remains an open problem for future research.

From the explicit construction of an $m$-forward pair in Theorem \ref{thm:mForwardConstructionHeterogeneous}, we obtain the following result showing that the forward utility $U_m$ depends in a measurable manner on the parameters of the financial market. 
This measurable dependence is crucial because it allows us to extend all results derived for an $m$-forward pair back to the level of a discrete-time predictable forward performance process with evaluation period length $m$.
 
\begin{corollary}\label{cor:Measurability_Heterogeneous}
Let $U_0 \in \mathcal{U}$ be a utility function and let
\begin{align}
    \mathcal{M} := \left\{ (p,u,d) \in \mathbb{R}^{m\times3} \big\vert 0 < p_i < 1, 0 < d_i < 1 < u_i, \enskip \{(\alpha_1,\dots,\alpha_i)\}_{i=1,\dots,m} \enskip \mathrm{exists} \right\}
\end{align}
be the set of market parameters under which $\{(\alpha_1,\dots,\alpha_i)\}_{i=1,\dots,m}$ exists and $p, u,$ and $d$ denote the $m\times1$ vectors $(p_i)_{i=1}^{m}$, $(u_i)_{i=1}^{m}$, and $(d_i)_{i=1}^{m}$ respectively. The mapping $\mathcal{M} \rightarrow \mathbb{R}$ defined by $(p,u,d) \mapsto U_m(x)$, where $U_m(x)$ is defined as in Theorem \ref{thm:mForwardConstructionHeterogeneous}, is Borel-measurable for any $x > 0$. 
\end{corollary}

\begin{remark}\label{rem:DecreasingIn-m}
We remark that $m$-forward processes, when they exist, are naturally decreasing in the evaluation period. 
This is a direct consequence from the fact that putting all one's wealth into the risk-free asset is an admissible strategy together with the martingale optimality principle satisfied by the $m$-forward process. 

\end{remark}

Having established an explicit construction of an $m$-forward process, we next provide a comparison between the discrete-time predictable forward performance process with evaluation period length $m$ and the single-period discrete-time forward process after $m$-periods.
We denote the latter process by $\tilde{U} = (\tilde{U}_k)_{k\in \mathbb{N}_0}$, the optimal wealth process corresponding to $\tilde U$ by $\tilde X$. We are interested in comparing $\tilde{U}_m$ with $U_m$. 
Given an initial performance criterion $U_0$ and the market parameters $(p,u,d)\in \mathbb{R}^{3m}$, the process  $\{\tilde{U}_1, \tilde{U}_2,\dots,\tilde U_m\}$ is constructed according to the general scheme outlined in Section 7 of \cite{angoshtari2020predictable}.   


\begin{proposition}\label{prop:Connection}
If the sequence $\{(\alpha_1,\dots,\alpha_i)\}_{i=1,\dots,m}$ exists, then the single-period forward process $\tilde U_i$ exists for $i =1, \dots, m$, and satisfies $\tilde{U}_m (x)=U_m (x)$ and $\tilde X_m^{*}(x)=X_m^{*}(x)$ for all $x > 0$. 
\end{proposition}

Let us emphasize that Proposition \ref{prop:Connection} does not imply that the single-period forward process generally coincides with the $m$-forward. 
In the general setting of single-period forward processes, market parameters are allowed to be updated period-by-period and only required to be predictable; {therein it does not even make sense to define the $m$-period forward utility.} 
But in the setting of this paper, market parameters are known across evaluation periods and, thus, the single-period forward utility is $m$-period ahead predictable, i.e., $\tilde{U}_m\in \mathcal{F}_{0}$.
Only if this is the case, the single-period forward process coincides with the $m$-forward at the end of the evaluation period. 

According to Proposition \ref{prop:Connection}, the optimal strategies of the single-period and $m$-forward coincide, even if market parameters are heterogeneous. 
This implies that the optimal strategies corresponding to an $m$-forward process are inherently myopic. 
They do not make use of the knowledge about future market parameters, even though this information is available.
This myopic behavior of optimal strategies constitutes a fundamental difference to the classical, backward expected utility setting. 
In our setting where market parameters are known across the evaluation period, maximizing expected utility from terminal wealth typically leads to optimal strategies that, at an intermediate time, depend on the market parameter values at future times. 
We will further discuss this difference between the forward and backward setting at the end of this section in Example \ref{ex:SumOf2Power_Homogeneous}.
We also state the hypothesis that the myopic behavior of optimal strategies corresponding to predictable forward processes is not specific to the binomial tree model of this paper but also holds for more general models studied in \citet{strub2021evolution} and \citet{angoshtari2022predictable}.

\subsection{The homogeneous case}

Allowing for heterogeneous market comes at the expense of quite involved notation and formulas. 
To build more insights, we will next consider the special case where model parameters are time-homogeneous across the given evaluation period. 
This will lead to considerably simplified expressions. 
Specifically, while market parameters are still updated at the beginning of each evaluation period, we assume that they are constant throughout each evaluation period, i.e., $p_i = p$, $u_i = u$, and $d_i = d$ for $i = 1, \dots m$.
Accordingly, we set $a=\frac{1-p}{p}$ $\frac{q}{1-q}$, $b=\frac{1-q}{q}$, and $c=\frac{1-p}{1-q}$. The price process of the risky asset in a given evaluation period corresponds to an $m$-period binomial tree with homogeneous coefficients: There are $m+1$ possible outcomes which, when ordered from the lowest price level to the highest, occur with the probabilities $\binom{m}{i}p^i(1-p)^{m-i}$, $i = 0,1,...,m$, where the transition probability is denoted by $p$ and $\binom{m}{i}=\frac{m!}{i!(m-i)!}$ are binomial coefficients. 
Therefore, \eqref{eq:FunctionalEquation} in this case can be written as
\begin{align}\label{eq:FctEQ_Multinomial}
I_0 (\hat{y}) = \sum\nolimits_{i=0}^{m} \binom{m}{i} q^i(1-q)^{m-i}I_m \left( \hat{y} \frac{q^i(1-q)^{m-i}}{p^i(1-p)^{m-i}} \right), \quad \hat{y} > 0.
\end{align}
Next, we characterize solutions to the linear functional equation \eqref{eq:FctEQ_Multinomial} in the class of inverse marginal functions and provide conditions for uniqueness. 

\medskip    
    
For a given initial utility function $U_0 \in \mathcal{U}$ and associated inverse marginal function $I_0 \in \mathcal{I}$, the above auxiliary functions (\ref{eq:DefPhiPsiZero_heter}) and (\ref{eq:DefPhiPsi_heter}) reduce to
\begin{align}\label{eq:DefPhiPsiZero}
\Phi_0^{0} (y) = I_0(acy) - bI_0(cy)  \quad \mathrm{and} \quad \Psi_0^{0} (y) = y^{-\rm{log}_ab}I_0(cy), \quad y > 0 ,
\end{align}
and
\begin{align}\label{eq:DefPhiPsi}
\begin{split}
    \Phi_i^{j}(y) & = \frac{(1+b)^i}{b^{j}}\bigg( \sum_{n_1=0,...,n_i=0}^{\infty}(-1)^{p_{(n_1, \dots n_i )}} b^{q_{j;(n_1, \dots n_i )}} I_0 \left( a^{r_{j;(n_1, \dots n_i )} +1} c^{i+1} y \right)\\ & \hspace{2.5cm} - b \sum_{n_1=0,...,n_i=0}^{\infty}(-1)^{p_{(n_1,\dots,n_i)}} b^{q_{j;(n_1,\dots,n_i)}} I_0 \left( a^{r_{j;(n_1,\dots,n_i)}} c^{i+1} y \right) \bigg),\\
    \Psi_i^{j}(y) & = y^{-{\rm{log}}_a b} \frac{(1+b)^i}{b^{j}} \sum\nolimits_{n_1=0,...,n_i=0}^{\infty}(-1)^{p_{(n_1,\dots,n_i)}} b^{q_{j;(n_1,\dots,n_i)}} I_0 \left( a^{r_{j;(n_1,\dots,n_i)}} c^{i+1} y \right),
    \end{split}
\end{align}
for $y > 0$, $i = 1, \dots , m-1$ and $j =0, 1 ,\dots, i$, where the exponents are defined as $p_{(n_1,\dots,n_i)} = \sum\nolimits_{k=1}^{i} n_k$,  $q_{j;(n_1,\dots,n_i)} = -\sum\nolimits_{k=1}^{j} n_k + \sum\nolimits_{k=j+1}^{i} n_k$,  and $r_{j;(n_1,\dots,n_i)} = \sum\nolimits_{k=1}^{j} n_k - \sum\nolimits_{k=j+1}^{i} (n_k+1)$.
We next define the sequence $(A_i)_{i=0,\dots,m}$, which will play a similar role as $\{(\alpha_1,\dots,\alpha_i)\}_{i=1,\dots,m}$, start by letting $A_0 = 0$ and then iteratively set
\begin{align*}
    A_{i+1} = \left\{ \begin{array}{ll}
        A_i + 1 & \mathrm{if} \enskip \left( \Phi_{i}^{A_{i}}, \Psi_{i}^{A_{i}} \right) \enskip \mathrm{satisfies} \enskip (C1), \\
     A_i    & \mathrm{if} \enskip \left( \Phi_{i}^{A_{i}}, \Psi_{i}^{A_{i}} \right) \enskip \mathrm{satisfies} \enskip (C2),
    \end{array} \right.
\end{align*}
for $i= 0, \dots m-1$. 
Here in the homogeneous market, we only need to count how many times is either $(C1)$ or $(C2)$ is satisfied, i.e. only the cumulative number of times where the functional pair $(\Phi,\Psi)$ is in case $(C1)$ respectively $(C2)$ matters, the location or order where each condition is satisfied is irrelevant. This is in contrast to the more general case of heterogenous market parameters, where also the instance in time where the pair satisfies either of the conditions needs to be recorded and influences the definition of the subsequent functional pair.

The following corollary follows directly from Theorem \ref{thm:mForwardConstructionHeterogeneous}.
It yields a \textit{construction method} of the homogeneous setting for an $m$-forward pair and presents an explicit relationship between the associated inverse marginal functions, $I_0$ and $I_m$. 


\begin{corollary}\label{cor:mForwardConstruction}
Let $U_0 \in \mathcal{U}$ be a utility function with associated inverse marginal function $I_0$ and suppose that $(A_i)_{i=0,\dots, m}$ exists. Define $I_m: (0,\infty) \rightarrow (0,\infty)$  by
\begin{align}\label{eq:ForwardInverseMarginal_Homogeneous}
    I_m(y) := \frac{(1+b)^m}{b^{A_m}}\sum\limits_{n_1=0,...,n_m=0}^{\infty}(-1)^{p_{(n_1,\dots,n_m)}}b^{q_{A_m;(n_1,\dots,n_m)}}I_0(a^{r_{A_m;(n_1,\dots,n_m)}}c^my), \quad y>0,
\end{align}
and $U_m: (0,\infty) \rightarrow (0,\infty)$ by
\begin{align*}
    U_m(x):=U_0(1)+ \mathbb{E}_{\mathbb{P}}\left[ \int_{I_m \left(\frac{d\mathbb{Q}}{d \mathbb{P}} U_0'(1)\right)}^{x} I_m^{-1}(t)dt\right], \quad x>0 .
\end{align*}
Then $U_m$ is the unique utility function solving the inverse investment problem \eqref{eq:InverseInvestmentProblem} and $I_m$ is the unique inverse marginal function  solving the linear functional equation \eqref{eq:FctEQ_Multinomial}. 
Moreover, the optimal wealth solving \eqref{eq:InverseInvestmentProblem} is given by 
\begin{align*}
X_m^*(x)=I_m \left( U_0'(x) \frac{d\mathbb{Q}}{d \mathbb{P}} \right). 
\end{align*}
\end{corollary}

\subsection{Two examples}
We close this section with two examples. 
The first example with a power utility function generalizes Corollary 6.4 in \cite{angoshtari2020predictable}. 
In the second example, the initial utility function corresponds to an inverse marginal function that is a sum of two power functions, see also \cite{he2021forward} and \cite{geng2017temporal}.

\begin{example}\label{ex:Power_Heterogeneous}
Let $U_0(x)=(1-\frac{1}{\theta})^{-1}x^{1-\frac{1}{\theta}}, x>0$, and assume that $1\neq \theta>0, \theta \neq-{\rm{log}}_ab$. By applying Theorem \ref{thm:mForwardConstructionHeterogeneous}, the unique inverse marginal function and the associated utility function in the case of heterogeneous market are given by  
\begin{align*}
    &I_m(x)=\prod \limits_{i=1}^m\left(\frac{1+b_i}{c_i^{\theta}(a_i^{-\theta}+b_i)}\right)y^{-\theta},\\
    &U_m(x)=\prod \limits_{i=1}^m\left(\frac{1+b_i}{c_i^{\theta}(a_i^{-\theta}+b_i)}\right)^{1/\theta}(1-\frac{1}{\theta})^{-1}x^{1-\frac{1}{\theta}}=\prod \limits_{i=1}^m\left(\frac{1+b_i}{c_i^{\theta}(a_i^{-\theta}+b_i)}\right)^{1/\theta}U_0(x).
\end{align*}
The corresponding optimal investment policy is given by 
\begin{align*}
    \pi_i=\frac{x}{u_i-d_i}\prod\limits_{k=1}^{i}\delta_kP_{(1,\dots,i-1)}\left((\frac{p_i}{q_i})^{\theta}-(\frac{1-p_i}{1-q_i})^{\theta}\right),
\end{align*}
where $P_{(1,\dots,i-1)}=\prod\limits_{j=1}^{i-1}\left((\frac{p_j}{q_j})^{\theta}\mathbbm{1}_{\{S_j > S_{j-1}\}}+(\frac{1-p_j}{1-q_j})^{\theta}\mathbbm{1}_{\{S_j < S_{j-1}\}}\right)$, and $\delta_k=\frac{1+b_k}{c_k^{\theta}(a_k^{-\theta}+b_k)}.$
\end{example}

\begin{example}\label{ex:SumOf2Power_Homogeneous}
Let $U_0(x)=\frac{2^{\frac{1}{\theta}-1}}{(1-\frac{1}{\theta})(2-\frac{1}{\theta})}\left(\sqrt{4x+1}-1\right)^{1-\frac{1}{\theta}}\left((1-\frac{1}{\theta})\sqrt{4x+1}+1\right)$, $x>0$, and assume that $\theta \neq 1$ and $\theta>\max \limits_{i=1,\dots,m}-\rm{log}_{a_i}b_i$ or $\theta<\min \limits_{i=1,\dots,m}-\frac{1}{2}\rm{log}_{a_i}b_i$. 
The marginal and inverse marginal of the initial datum can be respectively computed by 
\begin{align*}
    U_0^{'}(x)&=\left(\frac{\sqrt{4x+1}-1}{2}\right)^{-\frac{1}{\theta}}, \quad x > 0\\
    I_0(y)&=(U_0^{'})^{-1}(y)=y^{-\theta}+y^{-2\theta}, \quad y > 0 ,
\end{align*}
One can easily argue that the sequence $\{ (\alpha_1 , \dots , \alpha_i )\}_{i = 1 ,\dots, m}$ exists under the assumption on $\theta$ and that $a_i\neq1, i=1,\dots,m$. Indeed, $\left( \Phi_{i}^{(\alpha_1,\dots,\alpha_i)}, \Psi_{i}^{(\alpha_1,\dots,\alpha_i)} \right)$ satisfies $(C1)$ when $a_i>1$ and $\theta>-\rm{log}_{a_i}b_i$ or $a_i<1$ and $\theta<-\frac{1}{2}\rm{log}_{a_i}b_i$, $\left( \Phi_{i}^{(\alpha_1,\dots,\alpha_i)}, \Psi_{i}^{(\alpha_1,\dots,\alpha_i)} \right)$  satisfies $(C2)$ when $a_i>1$ and $\theta>-\rm{log}_{a_i}b_i$ or $a_i<1$ and $\theta<-\frac{1}{2}\rm{log}_{a_i}b_i$. Moreover, we note that the sequence $\{ (\alpha_1 , \dots , \alpha_i )\}_{i = 1 ,\dots, m}$ always exists for any $1\neq \theta>0$ if $\rm{log}_{a_i}b_i>0,$ $i=1,\dots,m$.
By Theorem \ref{thm:mForwardConstructionHeterogeneous}, the unique $m$-forward inverse marginal function in the case of heterogeneous market is then given by
\begin{align*}
    &I_m(y)=\prod\limits_{j=1}^{m}\delta^{(1)}_jy^{-\theta}+\prod\limits_{j=1}^{m}\delta^{(2)}_jy^{-2\theta},
\end{align*}
where $\delta^{(1)}_j= \frac{1+b_j}{c_j^\theta(a_j^{-\theta}+b_j)}, \delta^{(2)}_j= \frac{1+b_j}{c_j^{2\theta}(a_j^{-2\theta}+b_j)}$, and the corresponding unique $m$-forward utility function is given by
{
\begin{align*}
    U_m(x) & = \frac{2^{\frac{1}{\theta}-1}
    \prod\limits_{j=1}^{m}(\delta^{(2)}_j)^{\frac{1}{\theta}-1}}{(1-\frac{1}{\theta})(2-\frac{1}{\theta})}\left(\sqrt{4\prod\limits_{j=1}^{m}\delta^{(2)}_jx + \prod\limits_{j=1}^{m}(\delta^{(1)}_j)^2} -\prod\limits_{j=1}^{m}\delta^{(1)}_j\right)^{1-\frac{1}{\theta}}\\
    & \qquad \times \left( \left(1-\frac{1}{\theta} \right) \sqrt{4\prod\limits_{j=1}^{m}\delta^{(2)}_jx + \prod\limits_{j=1}^{m}(\delta^{(1)}_j)^2} +\prod\limits_{j=1}^{m}\delta^{(1)}_j\right),
\end{align*}}
with the optimal investment policy given by
\begin{align*}
    \pi_i=\frac{\sqrt{4x+1}-1}{2(u_i-d_i)}&\left(\prod\limits_{j=1}^{m}\delta^{(1)}_jP_{(1,\dots,i-1)}\left((\frac{p_i}{q_i})^{\theta}-(\frac{1-p_i}{1-q_i})^{\theta}\right)\right.\\&\left.\qquad+\left(\frac{\sqrt{4x+1}-1}{2}\right)\prod\limits_{j=1}^{m}\delta^{(2)}_jP_{(1,\dots,i-1)}^{(2)}\left((\frac{p_i}{q_i})^{2\theta}-(\frac{1-p_i}{1-q_i})^{2\theta}\right)\right),
\end{align*}
where $P_{(1,\dots,i-1)}^{(2)}=\prod\limits_{j=1}^{i-1}\left((\frac{p_j}{q_j})^{2\theta}\mathbbm{1}_{\{S_j > S_{j-1}\}}+(\frac{1-p_j}{1-q_j})^{2\theta}\mathbbm{1}_{\{S_j < S_{j-1}\}}\right)$.
\end{example}
\begin{remark}
We further illustrate the independence of optimal strategies on future parameters. 
For this sake, we compute the $m$-forward utility and the corresponding optimal investment under a market setup consisting of two trading periods,  note that here we only consider the case of $m=2$ to avoid complicated permutations of model parameters. 
We denote the market parameters by $\{(p_i, u_i, d_i)\}_{i=1}^2$ and the $2$-forward inverse marginal by $I_2(y)$.
We further set market parameters such that $q_1=q_2=q<p_1<p_2$, where $q_i=\frac{1-d_i}{u_i-d_i}, i=1,2$ are the associated risk-neutral probabilities for period $[i-1,i)$.
This implies a worse market for the first period and a better market for the second period. 

Let $\delta_1^{(1)}=\frac{1}{p_1^{\theta}q^{1-\theta}+(1-p_1)^{\theta}(1-q)^{1-\theta}}$, $\delta_1^{(2)}=\frac{1}{p_1^{2\theta}q^{1-2\theta}+(1-p_1)^{2\theta}(1-q)^{1-2\theta}}$, $\delta_2^{(1)}=\frac{1}{p_2^{\theta}q^{1-\theta}+(1-p_2)^{\theta}(1-q)^{1-\theta}}$, and $\delta_2^{(2)}=\frac{1}{p_2^{2\theta}q^{1-2\theta}+(1-p_2)^{2\theta}(1-q)^{1-2\theta}}$.
By Theorem \ref{thm:mForwardConstructionHeterogeneous}, one can derive that $I_2(y)=\delta_1^{(1)}\delta_2^{(1)}y^{-\theta}+\delta_1^{(2)}\delta_2^{(2)}y^{-2\theta}$, and the optimal strategy for trading period $[0,1)$ is given by
\begin{align*}
    \pi_1^{*}(x)=\frac{1}{u_1-d_1}&\left((U_0^{'}(x))^{-\theta}\delta_1^{(1)}\left((\frac{p_1}{q})^{\theta}-(\frac{1-p_1}{1-q})^{\theta}\right)+(U_0^{'}(x))^{-2\theta}\delta_1^{(2)}\left((\frac{p_1}{q})^{2\theta}-(\frac{1-p_1}{1-q})^{2\theta}\right)\right),
\end{align*}
Conforming with the implications of Proposition \ref{prop:Connection}, the optimal strategy depends only on the market information of the first period and do not make use of the information about future market parameters. 

To further highlight the myopic nature of forward optimal policies, we compare it with the classical, backward setting considering $U_0$ as a utility from terminal wealth.
The associated inverse marginal at time $t=2$ is given by $I_0(y) =y^{-\theta}+y^{-2\theta}$. 
We readily obtain the optimal strategy for the backward setting and find that 
\begin{align*}
    &\pi_1^{*}(x) = \frac{1}{u_1-d_1}\left((U_0^{'}(x))^{-\theta}\frac{1}{\delta_2^{(1)}}\left((\frac{p_1}{q})^{\theta}-(\frac{1-p_1}{1-q})^{\theta}\right)+(U_0^{'}(x))^{-2\theta}\frac{1}{\delta_2^{(2)}}\left((\frac{p_1}{q})^{2\theta}-(\frac{1-p_1}{1-q})^{2\theta}\right)\right),
\end{align*}
We observe that $\pi_1^{*}(x)$ then depends on the associated future parameters, $\delta_2^{(1)}$ and $\delta_2^{(2)}$, of the second period.
\end{remark}

\section{Application to automated trading}\label{sec:AutomatedTrading}
Trading is increasingly automated and contains only little human involvement and oversight. 
For example, as documented in 
Bloomberg\footnotemark, algorithmic trading accounts for around 60-73$\%$ of
all U.S. equity trading.
Much of the literature has focused on the aggregated
market impact of automated trading and the design of such computer
based strategies with the aim of minimizing the overall market impact cost, reducing transaction cost and the exposure
to timing risk, and self-adjusting to current market conditions (see, e.g., \cite{chaboud2014rise}, \cite{menkveld2013high}, 
and \cite{fabozzi2010quantitative}). 
Herein, we cover an aspect of automated trading that is new to the best of our knowledge: 
We study a setting where a human expert periodically updates her assessment of market parameters as inputs for a trading algorithm. 
In between those updates, trading is automated. 
The key question we aim to study is how frequently interaction between human expert and machine trader should occur when assessing the market is costly for the expert. 


We consider a human expert operating an automated trading system (ATS).
The expert periodically assesses the market and updates market parameters as inputs to the ATS. 
After an initial input of preferences and model for the market, the ATS trades automatically on behalf of and in the best interests of the expert until the next interaction time with the expert. 
At each interaction, the expert updates her assessment of the market and communicates the updated market parameters to the ATS, which then continues investing on behalf of the expert until the next point of contact. This procedure goes on indefinitely. 
Gathering information and assessing the market are costly, and the expert thus faces a trade-off between the ATS operating on an outdated model for the market and the costs associated with frequently updating the model. 

The setting outlined above can formally be described as follows. 
At time zero, the expert communicates her initial preferences in terms of a utility function $U_0$, her assessment of market parameters $(p_1,u_1,d_1)$, and the intended interaction schedule $m \in \mathbb{N}$. 
The ATS then determines the $m$-forward according to the construction we provided in Section \ref{sec:FunctionalEquation} and trades period-by-period on behalf of the expert assuming that the model parameters for the entire investment period up to time $m$ is given by $(p_1,u_1,d_1)$. 

The expert reassesses the market whenever she interacts with the ATS. 
This results in new, accurate model parameters $(p,u,d)$ but also imposes a cost to the expert reflecting the time and effort to newly calibrate the model and interacting with the ATS. 
We herein suppose that $u$ and $d$ do not vary over time, and the expert thus only needs to reassess the probability $p$ for an upward movement of the stock.
We further suppose that the cost is proportional to the expert's current wealth in the following sense: 
Whenever the expert interacts with the ATS, her wealth is reduced from $x$ to $\alpha x$ for some $0 < \alpha \leq 1$. 

On the one hand, when the expert interacts with the ATS infrequently, she saves on the interaction cost but faces the risk that the ATS trades based on a model with outdated  market parameters. 
On the other hand, when the expert interacts with the ATS very frequently, the ATS will trade based on an accurate model for the market but the expert incurs a heavy interaction cost. 
The expert thus attempts to balance a tradeoff between interaction cost and accuracy of the model parameters serving as input to the ATS. 
In order to evaluate the performance of alternative interaction schedules, we consider an investment horizon $T$ and denote by $\mathbb{P}^{(1)}$ the probability measure where market parameters are updated in each period. 
The operator $\mathbb{E}^{(1)}$  denotes the expectation under $\mathbb{P}^{(1)}$ and by $U^{(1)}$ the $1$-forward process. 
The performance of an interaction schedule corresponding to an evaluation period length $m<T$ is given by $\mathbb{E}^{(1)} \left[ U^{(1)}_T (X^{(m)}_T) \right]$, where optimal wealth $X^{(m)}$ corresponds to the $m$-forward process after transaction costs.

In the following, we study two approaches of determining the optimal interaction schedule $m^{*}$: a robust approach and a specific example where the updating rule is described by a maximum likelihood estimator.
We focus on an initial datum belonging to the CRRA family of utility functions. 
On the one hand, this leads to more tractable results. 
We know from our earlier discussion in Section \ref{sec:FunctionalEquation} that Theorem \ref{thm:mForwardConstructionHeterogeneous} applies unless $\theta = - \log_{a} b$. In this special case,  all we lose is uniqueness of the solution and we can in particular still apply Theorem \ref{thm:mForwardConstructionHeterogeneous}.

\footnotetext{https://www.bloomberg.com/press-releases/2019-02-05/global-algorithmic-trading-market-to-surpass-us-21-685-53-million-by-2026, accessed on 4th Feb, 2023.}

\subsection{Optimal interaction schedule under robust approach}
\label{subsec:ATS_Robust}

We first consider a robust approach and seek to derive bounds on $\mathbb{E}^{(1)} \left[ U^{(1)}_T (X^{(m)}_T) \right]$ which hold under any predictable updating rule within a certain class. 
Specifically, we consider  transition probabilities given in reference to the initial probability $p$ by $p_i=D_ip$, where $D_i$ is a $\mathcal{F}_{i-1}$-measurable random variable and can take any value in the interval $[D_{i,d}, D_{i,u}]$. 
The interval bounds $D_{i,d}$ and $D_{i,u}$ are some constants satisfying $0<D_{i,d}\leq 1 \leq D_{i,u}$ and depend on the choice of interaction schedule $m$. 
To maintain absence of arbitrage, we must have $0 < p_i < 1$, and it is thus without loss of generality that $D_i$ takes values in a bounded interval.

It seems plausible to assume that $D_{i,u}$ increases and $D_{i,u}$ decreases over time in any evaluation period $(km, (k+1)m]$, $k\in \mathbb{N}_{0}$, and then resets to a smaller level at the beginning of next period after a new interaction between human expert and the ATS.
Indeed, this reflects the intuition that,  as time since last calibrating the model passes, the expert becomes more uncertain about model parameters. 
We model this behavior by assuming periodicity on $D_{i,u}$ and $D_{i,d}$ in the interaction schedule $m$, i.e., denoting by $\mod$ the modulo operator,
$D_{i,d} = \tilde{D}_{({i \mod m}),d}$ and $D_{i,u}= \tilde{D}_{({i \mod m}), u}$, for $i=1,\dots,T$, for exogenously given sequences $(\tilde{D}_{i,d})_{i=1}^T$ and $(\tilde{D}_{i,u})_{i=1}^T$ satisfying that $\tilde{D}_{i,u} - \tilde{D}_{i,d}$ is increasing in $i = 1, \dots, T$.
We also assume that $\tilde D_{1,u}=\tilde D_{1,d}=1$, corresponding to accurate market parameters at any time there is an interaction. 

\begin{proposition}\label{prop:BoundsPerformanceSchedule}
Suppose that the initial datum is of the form $U_0(x)=(1-\frac{1}{\theta})^{-1}x^{1-\frac{1}{\theta}},$ $x>0$, for some $1 \neq \theta>0$, and let $T \in \mathbb{N}$ be an evaluation horizon. 
Let $m$ be an interaction schedule that is a divisor of $T$, i.e., $m \in \mathbb{N}$ and $T/m \in \mathbb{N}$, and let $(D_i)_{i=1,\dots, T}$ be a predictable process taking values in $[D_{i,d},D_{i,u}]$, where $D_{i,d}$ and $D_{i,u}$ satisfy the assumption of periodicity in the interaction schedule $m$ and are such that absence of arbitrage is maintained.
Then, the optimal expected performance value $\mathbb{E}^{(1)} \left[ U^{(1)}_T (X^{(m)}_T) \right]$ lies in
\begin{align*}
\left[\alpha^{(\frac{T}{m}-1)(1-\frac{1}{\theta})} \left(\prod\limits_{j=1}^{m}f_j \right)^{\frac{T}{m}}\delta^{T(1-\frac{1}{\theta})} U_0(x), \alpha^{(\frac{T}{m}-1)(1-\frac{1}{\theta})}U_0(x) \right],
\end{align*}
where $1-\alpha$ denotes the proportional interaction cost and $f_j$ are given by $f_{1}=\delta^{\frac{1}{\theta}-1}$ and $f_j={\rm{min}}\{f_{D_{j,u}},f_{D_{j,d}}\}$ if $\theta>1$, respectively $f_j={\rm{max}}\{f_{D_{j,u}},f_{D_{j,d}}\}$ if $\theta<1$, with
 \begin{align*}
     f_{D_{j,u}} &= \frac{C_1}{\left(C_1 + \left(\frac{D_{j,u}(1-p)}{1-D_{j,u}p} \right)^{-\theta}C_2\right)^{\frac{1}{\theta}}} + \frac{C_2}{\left( \left(\frac{D_{j,u}(1-p)}{1-D_{j,u}p} \right)^{\theta}C_1+C_2 \right)^{\frac{1}{\theta}}} ,\\
     f_{D_{j,d}} &= \frac{C_1}{\left( C_1+ \left(\frac{D_{j,d}(1-p)}{1-D_{j,d}p} \right)^{-\theta}C_2 \right)^{\frac{1}{\theta}}} + \frac{C_2}{\left( \left( \frac{D_{j,d}(1-p)}{1-D_{j,d}p} \right)^{\theta}C_1+C_2 \right)^{\frac{1}{\theta}}},
 \end{align*}
where $C_1=p^{\theta}q^{1-\theta}$, $C_2=(1-p)^{\theta}(1-q)^{1-\theta}$, and $j=2,\dots,m$. Furthermore, $f_{j}$ is non-increasing in $D_{j,u}$ and non-decreasing in $D_{j,d}$.
\end{proposition}

The following proposition shows that the optimal interaction schedule $m^*$ is independent of the evaluation horizon $T$ when the trader takes a robust approach of maximizing the minimal expected performance and 
the sequence of intervals $[D_{i,d}, D_{i,u}], i=1,2,..., T$ is periodic as assumed above.


\begin{proposition}
\label{prop:IndependentT}
Suppose that the initial datum is of the form $U_0(x)=(1-\frac{1}{\theta})^{-1}x^{1-\frac{1}{\theta}},$ $x>0$, for some $1 \neq \theta>0$. Let $T \in \mathbb{N}$ be an evaluation horizon, and let $(D_{i,d})_{i=1}^{\infty}$ and $(D_{i,u})_{i=1}^{\infty}$ be periodic in the interaction schedule and such that absence of arbitrage is maintained. 
There exists an optimal updating schedule $m^{*}$ maximising the minimal expected performance for any $T$ which is a multiple of $m^{*}$. 
Moreover, the optimal interaction schedule $m^*$ can be determined by maximising the function $\mathbb{N} \rightarrow \mathbb{R}$ given by
\begin{align}\label{eq:Coefficient_MinExpectedPerformance}
    m \mapsto    \left(\alpha^{1-\frac{1}{\theta}} \prod\limits_{j=1}^m (f_j \delta^{1-\frac{1}{\theta}}) \right)^{\frac{1}{m}}.
\end{align} 
\end{proposition}

According to Proposition \ref{prop:IndependentT}, the optimal interaction schedule $m^*$  depends on 
the market parameters $p$, $u$, and $d$, the (constant) Arrow-Pratt measure of relative risk aversion $1/\theta$ of the initial datum of a trader, and the uncertainty about the evolution of future market performance captured in the sequences $D_{i,d}$ and $D_{i,u}$, but not on the evaluation horizon $T$. 
In practice, at time zero, the expert chooses $m^{*}$ based on the current understanding of the market. 
At the subsequent interaction time, market parameters are updated, and a new optimal interaction schedule is being chosen. 
Therefore, market parameters, updating frequencies, preferences and investment strategies move together forward in time. 

\medskip

Suppose for the following discussion that $\theta > 1$, the case where $\theta < 1$ can be treated similarly.
Since $f_j\delta^{1-\frac{1}{\theta}} \in (0,1]$, the term $\prod\limits_{j=1}^m (f_j\delta^{1-\frac{1}{\theta}})$ is decreasing in $m$. 
On the other hand, the base which belongs to $(0,1]$ raised to the power $1/m$ is increasing in $m$. 
Hence, there are two extreme cases: 
First, when the rate of decline in $\alpha^{1-\frac{1}{\theta}} \prod\limits_{j=1}^m (f_j \delta^{1-\frac{1}{\theta}})$ is very slow, i.e., when the probability for a positive return hardly varies over different periods, then \eqref{eq:Coefficient_MinExpectedPerformance} is strictly increasing in $m$. 
In this case, the strategy of never interacting is optimal, $m^*$ tends to $+ \infty$. 
Second, when the rate of decline in $\alpha^{1-\frac{1}{\theta}} \prod\limits_{j=1}^m (f_j \delta^{1-\frac{1}{\theta}})$ is very fast, i.e., when $p_i$ changes substantially across periods and the updating cost is small, then \eqref{eq:Coefficient_MinExpectedPerformance} is strictly decreasing in $m$. 
In this case, the strategy of period-by-period updating is optimal, i.e., $m^* = 1$. 
As we discussed earlier, the rate of decline is typically slow at first and then increases as more and more time elapsed since the last interaction time as a consequence of the increasing width $D_{i,u} - D_{i,d}$. Also, the rate of decline in $1/m$ is strictly decreasing, which implies that the degree of growth resulted from the decreasing exponent is weakening as $m$ increases. If this is the case, we are typically able to determine a unique optimal interaction time $m^*$ which is larger than one. 



\medskip

Intuitively, $m^*$ is increasing in the interaction cost and decreasing in the uncertainty about parameters. These are the two competing forces in our model, and $m^*$ attempts to find an ideal balance between them. 
In the following, we will confirm this intuition. We retain the assumption that the sequence of intervals $[D_{i,d}, D_{i,u}]$ is periodic and consider the case where $\theta > 1$. 

First, from the above analysis one can directly infer that $m^*$ is increasing in the interaction cost. Indeed, when $\alpha$ decreases, the rate of decline in $\alpha^{1-\frac{1}{\theta}} \prod\limits_{j=1}^m (f_j \delta^{1-\frac{1}{\theta}})$ is slower and the optimal $m^{*}$ that maximises \eqref{eq:Coefficient_MinExpectedPerformance} will thus be larger. This implies that the expert should interact less frequently with the ATS if interaction comes at a high cost.

Second, $m^{*}$ is typically decreasing in the uncertainty about parameters.
In other words, one should update more frequently when there is a larger range of possible values for the transition probability, while it is better to update less frequently when the parameter is stable and we can estimate it with more confidence.  
To substantiate this intuition, we consider a uniform increase of uncertainty and approximate it by the case where all factors $f_j, j=2,\dots,m$ simultaneously decrease to $f_j^{'}=Cf_j$, $j=2,\dots,m,$ with the same constant $C<1$, but $f_1^{'}=f_1$. 
The rate of decline in $\alpha^{1-\frac{1}{\theta}} \prod\limits_{j=1}^m (f_j^{'} \delta^{1-\frac{1}{\theta}})$ becomes quicker than $\alpha^{1-\frac{1}{\theta}} \prod\limits_{j=1}^m (f_j \delta^{1-\frac{1}{\theta}})$, hence $m^{*}$ that maximises \eqref{eq:Coefficient_MinExpectedPerformance} will be smaller. 
Moreover, consider any two alternative interaction schedules $m_1<m_2$ where $m_1$ outperforms $m_2$ before the increase in uncertainty, i.e., 
\begin{align}\label{eq:relation} \left(\alpha^{1-\frac{1}{\theta}}\prod\limits_{j=1}^{m_1}(f_j\delta^{1-\frac{1}{\theta}})\right)^{\frac{1}{m_1}} 
    > \left(\alpha^{1-\frac{1}{\theta}}\prod\limits_{j=1}^{m_2}(f_j\delta^{1-\frac{1}{\theta}})\right)^{\frac{1}{m_2}}.
\end{align}
Since $\frac{1}{m_1}>\frac{1}{m_2}$ and $C^{1-\frac{1}{m_1}}>C^{1-\frac{1}{m_2}}$, we then also have
\begin{align*}
    \left(\alpha^{1-\frac{1}{\theta}}C^{m_1-1}\prod\limits_{j=1}^{m_1}(f_j\delta^{1-\frac{1}{\theta}})\right)^{\frac{1}{m_1}}&=C^{1-\frac{1}{m_1}}\left(\alpha^{1-\frac{1}{\theta}}\prod\limits_{j=1}^{m_1}(f_j\delta^{1-\frac{1}{\theta}})\right)^{\frac{1}{m_1}}\\&>C^{1-\frac{1}{m_2}}\left(\alpha^{1-\frac{1}{\theta}}\prod\limits_{j=1}^{m_2}(f_j\delta^{1-\frac{1}{\theta}})\right)^{\frac{1}{m_2}}\\&=\left(\alpha^{1-\frac{1}{\theta}}C^{m_2-1}\prod\limits_{j=1}^{m_2}(f_j\delta^{1-\frac{1}{\theta}})\right)^{\frac{1}{m_2}},
\end{align*}
which means that an infrequent schedule $m_2$ cannot perform better than $m_1$ after uncertainty increases uniformly.

However, one needs to be more careful when the increase of uncertainty is not uniform. 
For example, suppose that only the $k$'th interval after every updating becomes wider, $D_{k,u}' - D_{k,d}' > D_{k,u} - D_{k,d}$, and all other parameters remain constant.
The original $f_k$ is reduced to a smaller $f_k^{'}$ by the last statement of Proposition \ref{prop:BoundsPerformanceSchedule}.
We investigate whether $m_2$ can outperform $m_1$ after the increase in three distinct cases: First, when $k>m_2$, \eqref{eq:relation} is not affected. Second, when $m_1<k \leq m_2$, the performance of schedule $m_1$ does not change, while the the performance of schedule $m_2$ decreases, and thus $m_1$ still leads to a better performance. 
However, in the third case: $k \leq m_1$, we might have the opposite inequality, i.e., $m_2$ outperforms $m_1$ after an increase in uncertainty. This happens because with lower interaction frequency, we can have less updating cost when its effect on the minimal expected performance is significant. 
More importantly, if $D_{k,u}^{'}$ becomes closer to $D_{k+1,u}$ such that the benefit one can get from the updating is small, then there is no need to update immediately. 
Therefore, we should not only just focus on the increasing loss incurred from deviation from the actual parameter, but also take the updating frequency and cost into consideration.

We close this section with two numerical examples. The market parameters of the first example for the table below are given by $p=0.6$, $u=1.2$, $d=0.8$, the constant relative risk aversion of the initial datum is $\theta=5$, the interaction cost is $0.2\%$, and the bounds for the updating of future probabilities are given by $D_{i,d}=D_d^{\mod(i-1,m)}, D_{i,u}=D_u^{\mod(i-1,m)}, i\in \{1,2,\dots,T\}$, where $D_d=0.99$, $D_u=1.01$.


\begin{table}[!htb]
	\centering
	\caption{Minimal expected performance at time $T$ for different interaction schedules}
	\begin{tabular}{*{12}{c}}
		\hline
		\multirow{2}{*}{$T=12$} & {$m$} & 1 & 2 & {\textcolor{red}{3}}& 4 & 6 & 12\\ \cline{2-8} 
		& {MEP} & $0.98$ & $0.991$ & ${\textcolor{red}{0.993}}$ & $0.991$ & $0.98$ & $0.93$\\ \hline \multirow{2}{*}{$T=24$} & {$m$} & 1 & 2 & {\textcolor{red}{3}} & 4 & 6 & 8 & 12 & 24\\ \cline{2-10} & {MEP} & $0.96$ & $0.981$ & ${\textcolor{red}{0.984}}$ & $0.981$ & $0.96$ & $0.94$ & $0.85$ & $0.48$\\ \hline \multirow{2}{*}{$T=48$} & {$m$} & 1 & 2 & {\textcolor{red}{3}} & 4 & 6 & 8 & 12 & 16 & 24 & 48\\ \cline{2-12} & {MEP} & $0.93$ & $0.96$ & ${\textcolor{red}{0.97}}$ & $0.96$ & $0.93$ & $0.88$ & $0.73$ & $0.55$ & $0.23$ & $0.003$\\ \hline
	\end{tabular}
\label{MatlabSimulation}
\end{table}

{\footnotesize{\textit{Notes}. The MEP (minimal expected performance) presented above is re-scaled by dividing it by $U_0(x)$. We only update at frequencies $1/m$ where $m$ is a divisor of $T$.}}

Conforming with Proposition \ref{prop:IndependentT}, Table \ref{MatlabSimulation} shows that the optimal interaction frequencies are independent of the evaluation horizon $T$.
While the minimal expected performance, MEP in the table, is decreasing over $T$ because losses in expected performance from both interaction and not updating timely are accumulating as time elapses, the optimal interaction schedule $m^{*}$ exists and is universal for any evaluation horizon $T$.
We can also infer that the minimal expected performance is first increasing and then decreasing with respect to $m$.
This demonstrates the tradeoff between updating cost and deviation from actual parameter due to not updating on the system in time. 

In the second example, we investigate the impact of increasing risk aversion on the optimal interaction schedule $m^{*}$. 
There are two distinct cases. 
Figure below visualizes how the optimal interaction schedule $m^{*}$ depends on the client's risk preference parameter $\theta>1$ and the difference between $p$ and $q$ when $D_d=0.99, D_u=1.01$. 

\begin{figure}[H]
   \includegraphics[width=7.96cm]{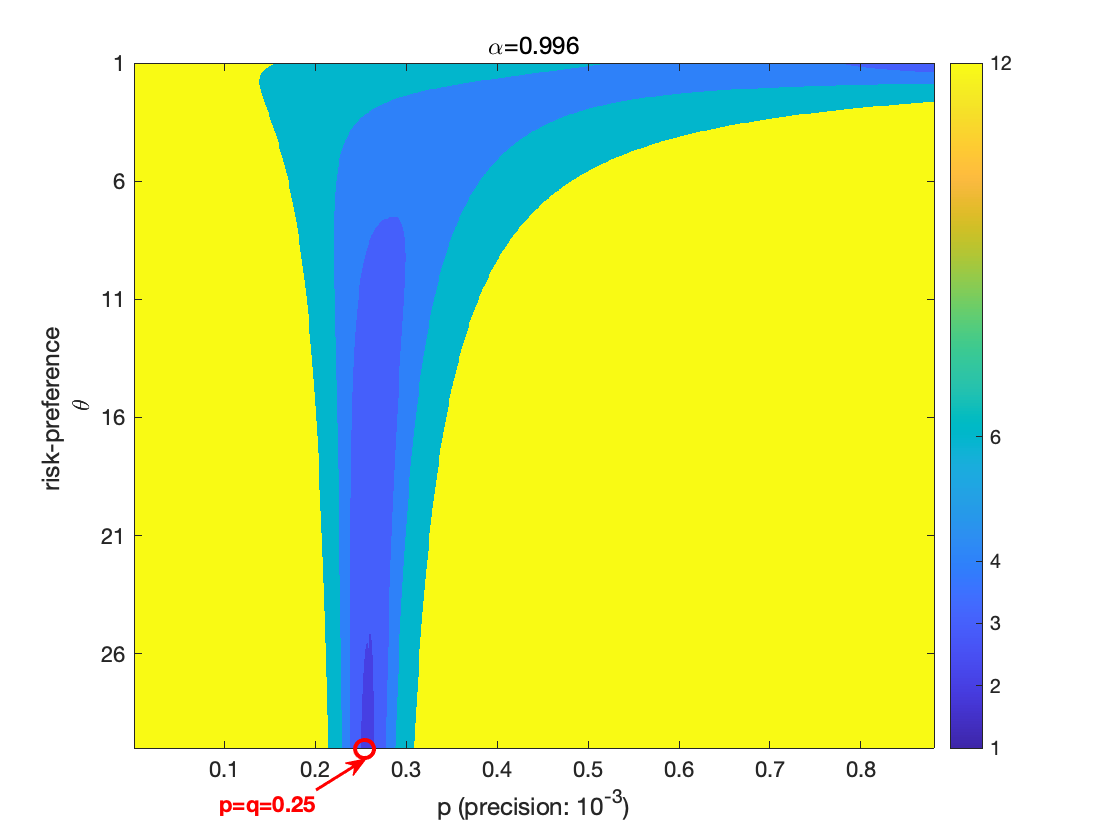}
  \includegraphics[width=7.96cm]{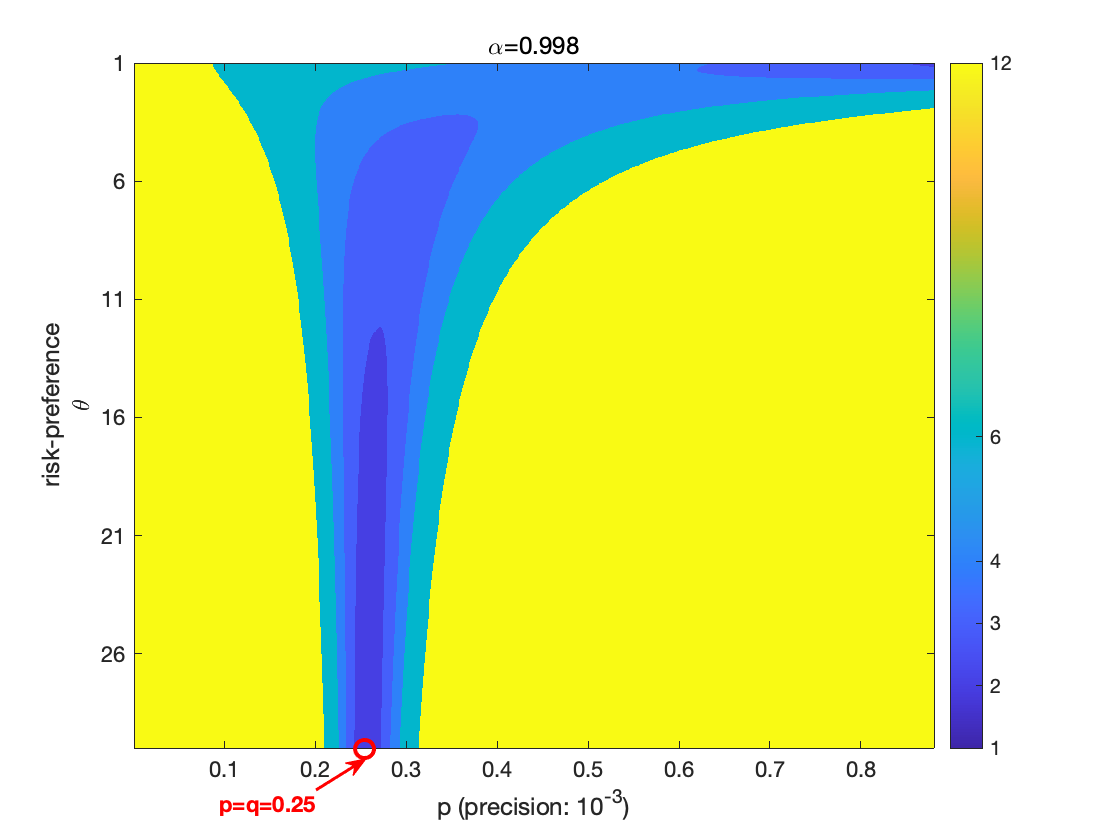}\\
  \includegraphics[width=7.96cm]{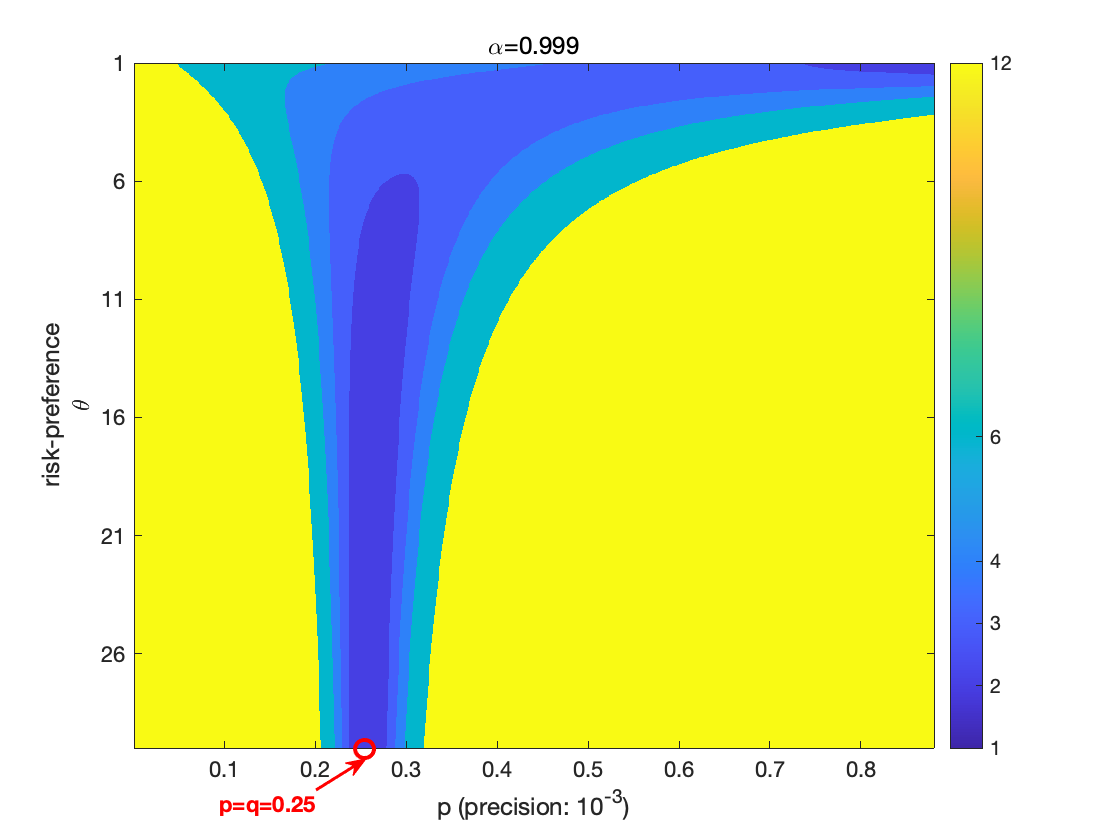}
  \includegraphics[width=7.96cm]{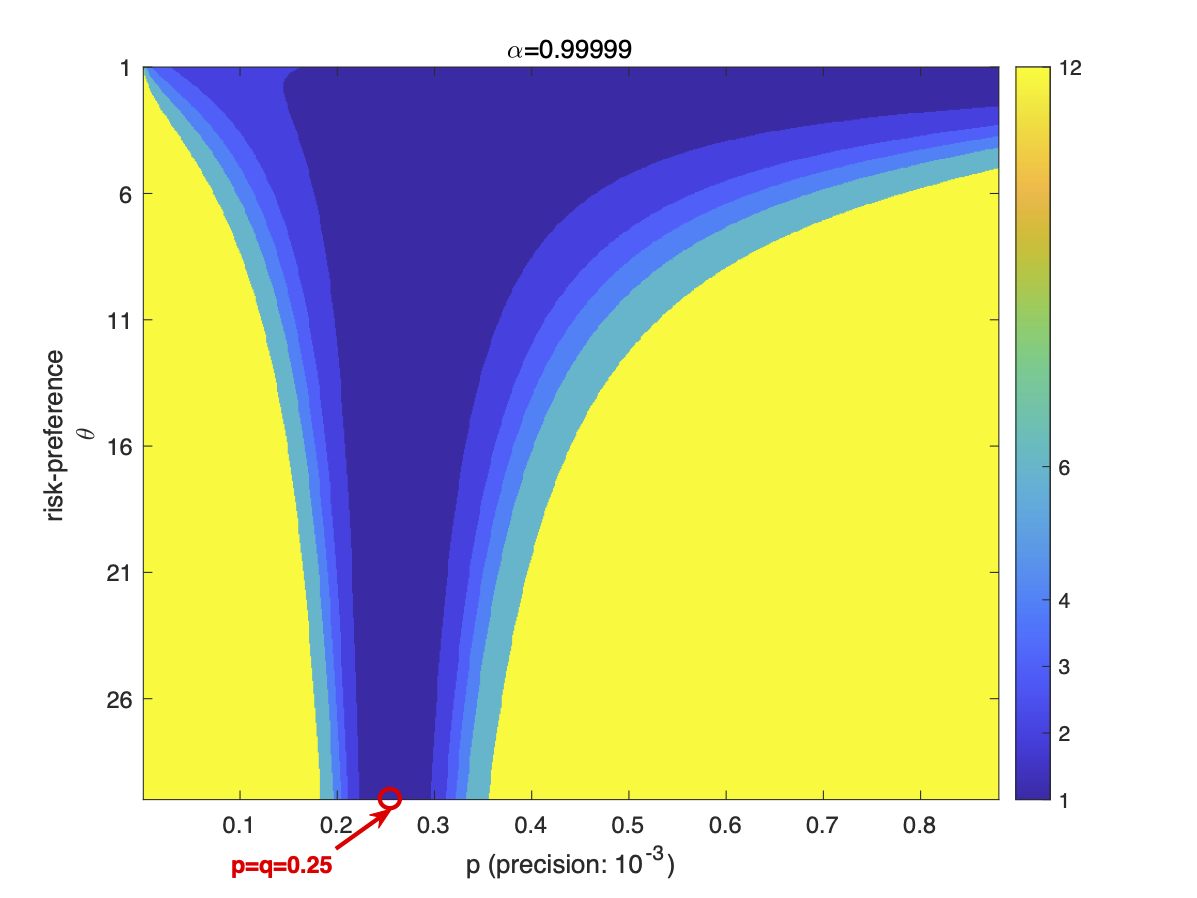}
  \caption{Optimal interaction schedule
computed with different market parameters.}
  \label{fig1}
\end{figure}

In the first case, when $|p-q|$ or $\theta$ is large enough, we observe  that a more risk averse expert is interacting more frequently with the ATS than a less risk averse expert.
However, in the second case where $p$ is close to $q$, or when the expert is already extremely risk-averse, we make the opposite observation that the expert decreases her interaction frequency as she becomes more risk-averse. 
This is because, in this case, the investment in the risky asset is very low, and the updating of the probability for a positive outcome does not lead to a significant change in optimal investment strategies.
This situation is especially likely to occur when, at the same time, the updating cost plays a relevantly important role in determining the optimal interaction schedule. 
Furthermore, we observe from all the first three heat maps that, as $\theta$ increases, or equivalently risk aversion decreases,
the region where the optimal interaction schedule increases as the expert becomes more risk averse becomes narrower around the region where $p = q$. 
The influence of the interaction cost on the width of this region depends on two competing forces. First, since the expert prefers interacting less frequently when faced with higher interaction cost as argued above, the set of $(\theta, p)$-combinations leading to an optimal interaction schedule $m^{*}$ that is smaller than the evaluation horizon (the areas of a color other than yellow in the heat map) are reduced to an increasingly narrow band around the value $p = q$ as interaction cost increases. 
However, the width might also become larger as the interaction cost grows in situations where the loss from each interaction outweighs the benefit of accurate knowledge about the model parameters. Therefore, the region where the optimal interaction schedule increases as the expert becomes more risk averse does not grow monotonically in the interaction cost as shown in Figure \ref{fig1}.

\subsection{Optimal interaction schedule for the maximum likelihood estimator for positive return probability}\label{MLEExample}


In this subsection, we study an explicit updating rule where the expert assesses probabilities for a positive return of the stock $p_t, t\in \mathbb{N}$ according to the maximum likelihood estimators given past information. By assuming that $p_{t+1} = \E[B_{t+1}|\mathcal{F}_t]$ for all $t$, where $(B_t)_{t \in \mathbb{N}}$ is the sequence of Bernoulli random variables associated with the stock price process, we are able to define the corresponding maximum likelihood estimators.
Specifically, suppose that there are $N$ observations about the past performances of risky asset at time zero, and that the stock achieves a positive return $N_u$ times.
The maximum likelihood estimator for an upward move of the stock in the first period [$0,1$) is thus given by $p_1=\frac{N_u}{N}$, in the second period [$1,2$)  by
\begin{align}
    p_2=\frac{Np_1+1}{N+1}\mathbbm{1}_{\left\{R_1=u\right\}}+\frac{Np_1}{N+1}\mathbbm{1}_{\left\{R_1=d\right\}}\label{5.1}
\end{align}
and so forth.
Let $N_t^u$ represent the process of total number of positive returns of the stock from time 0 until time $t$ starting from $N_0^u=0$. 
We then have for $t=1,2,3,...,m-1$,
\begin{align}
    p_{t+1}=\frac{Np_1+N_t^u}{N+t},\quad 1-p_{t+1}=\frac{N(1-p_1)+t-N_t^u}{N+t}\label{5.2} .
\end{align}

As in the previous section, we seek to determine an interaction schedule that represents an optimal trade-off between loss in performance value due to the deviation from the actual assessment of the market and the updating cost occurring whenever the expert assesses and communicates market parameters.  
We limit our analysis on a numerical example comparing two settings where the initial assessments of $p_1$ are identical, but one is based on a larger number of past observations than the other.
The parameter values for this example are $u=1.3$, $d=0.8$, $\theta=3$, $m\in\{1,2,3,4,6,12\}$ which are the factors of $T=12$. 
We again consider an initial utility function of the form $U_0(x)=(1-\frac{1}{\theta})^{-1}x^{1-\frac{1}{\theta}}$, $x>0$, a proportional interaction cost set to $\alpha = 0.4\%$, and suppose that the initial wealth is $X_0=9'960$ corresponding to an initial wealth of $10'000$ minus the interaction cost.
We perform $10^8$ simulations to compute all involved expected values.



\begin{figure}
    \centering
    \includegraphics[scale=0.31]{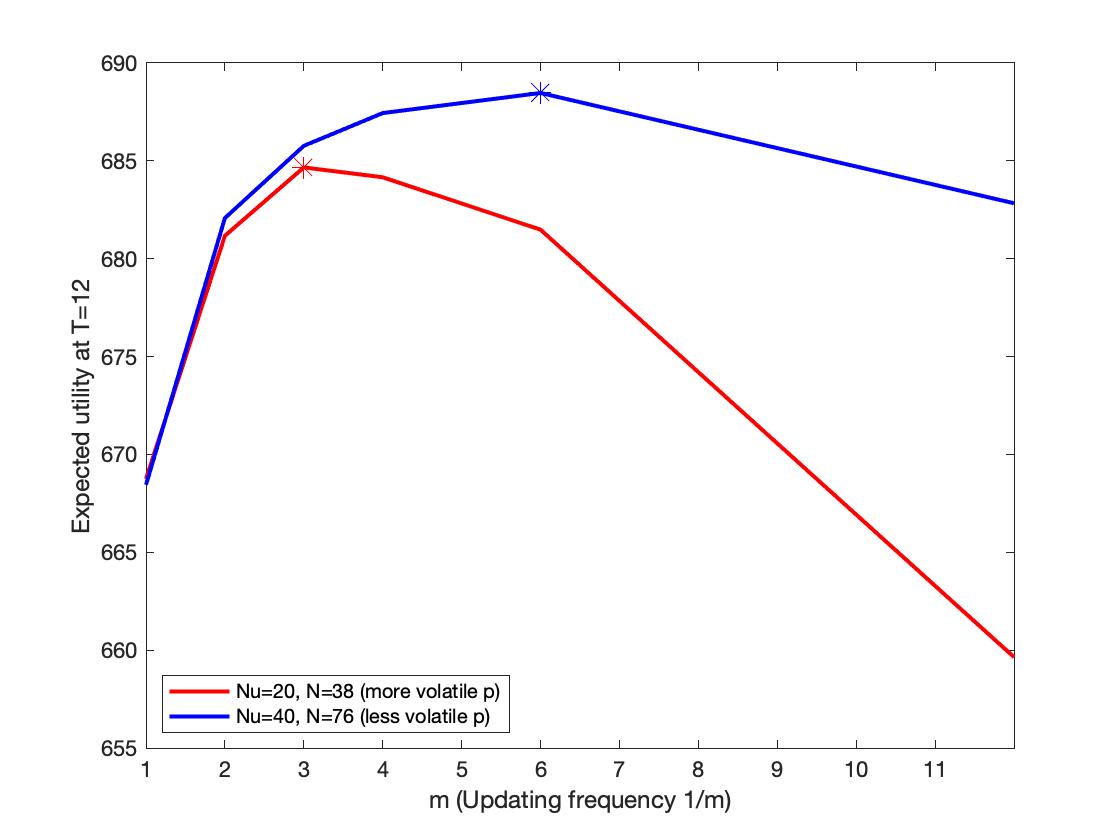}
    \caption{Numerical example of finding optimal updating frequency for terminal expected utility $\E_{1} [U_T^{(1)} ({X}^{(m)}_T)]$ at $T=12$.
    }
    \label{fig2}
\end{figure}

Figure \ref{fig2} shows the optimal interaction schedule $m^{*}$ at which the expected performance is maximal. 
We observe that the expected performance is first increasing and then decreasing as a function of the interaction schedule $m$.
This is what we expect: On the one hand, if the expert interacts with the ATS too frequently, the loss due to the interaction cost dominates. 
But, on the other hand, if there are too few interactions and updates in parameters are not communicated to the ATS in a timely manner, the loss due to inaccurate model parameters dominates.

The blue and red scenarios correspond to settings where we have more (blue), respectively less (red), prior observations of the stock performance.
When we have a large number of prior observations, our assessment of the probability of an upward movement is less susceptible to a single new observation than when we have fewer observations. 
This translates to a larger interaction schedule $m$ being optimal, since it becomes less important to immediately adjust to updated assessment of the market on the system.

\section{Robo-advising applications}\label{sec:RoboAdvising}

Personalized robo-advisors provide automatized advice on asset allocation and investment strategies. 
They provide wealth management services for large number of clients and at lower cost than traditional financial advisors.
Robo-advising companies constitute a rapidly growing part of the financial industry and are a prime example of FinTech, the application of technology to improve financial services. 
In this section, we propose and discuss discrete-time predictable forward performance processes as a potential framework for guiding asset allocation decisions of robo-advisors. 

\subsection{Preference modelling for robo-advising applications}\label{sec:RoboAdvising_Literature}

Although robo-advising has rapidly grown in popularity  over the last decade and now constitutes an important segment of modern investment industry, there is surprisingly little existing research on preference modeling for robo-advising applications and on the quantitative modelling of asset allocation decisions within those systems. \cite{capponi2022personalized} and \cite{cui2022risk} were the first papers discussing the portfolio optimization part of robo-advisors quantitatively. 
While \cite{capponi2022personalized}  proposed an adaptive mean-variance control model with updating of the risk aversion for deriving optimal allocation policies, \cite{cui2022risk} considered the framework of mean-variance induced utility functions and argued that this approach has several desirable features from the perspective of robo-advisors. 
A further important study is \cite{dai2021robo} who consider the mean-variance objective for log returns introduced in \cite{dai2021dynamic}, and provide an explicit formula for eliciting preferences in this setting. 
A comparison of the key features of asset allocation models for robo-advising is given in Table \ref{tab:Comparison}.

\begin{table}[!htb]
	\centering
	\caption{Comparison of main features with key literature}
	\begin{tabularx}{\textwidth}{l  >{\raggedright\setlength\hsize{1.7\hsize}}X >{\raggedright\setlength\hsize{.8\hsize}}X >{\setlength\hsize{.5\hsize}}X}
        \hline
        & Performance criterion & Investment horizon & Market model\\
         \hline
        \cite{capponi2022personalized} & Mean-variance with exogeneous updating of risk aversion & finite,\newline set ex ante &  discrete-time\\
        \cite{cui2022risk} & Mean-variance induced utility maximization &  finite,\newline set ex ante &  discrete-time \\ 
        \cite{dai2021robo} & Mean-variance for log returns &  finite,\newline set ex ante & continuous-time\\
        This paper & $m$-forward process, endogeneous updating of preferences & flexible & discrete-time\\
        \hline
    \end{tabularx}
    \label{tab:Comparison}
\end{table}
 
The work of \cite{capponi2022personalized} is most closely related to our paper and inspired many of the ideas we will subsequently discuss.
In their model, the market dynamics depends on an observable time-homogeneous Markov chain representing economic regimes. 
Preferences of the agent are modelled according to a multi-period mean-variance objective with a finite investment horizon. 
A key feature of their model is that the risk preferences of the agent are dynamic and stochastic. 
However, the robo-advisor cannot observe the risk preferences of the agent at all times and thus has to construct a proxy risk aversion process which is then used in the dynamic mean-variance optimziation problem. 
Only at times when the client and robo-advisor interact will the latter become aware of the idiosyncratic component of the client's risk preferences. 
Since interaction times occur at a slower pace than trading times, the robo-advisor has to automatically construct a proxy for the risk preferences of the agent and trade on her behalf between two consecutive interaction times. 
\medskip

The setting where trading occurs at a higher frequency than performance measurement updating is reminiscent of the framework of $m$-forward processes we study herein and thus prompted us to explore possible applications of our results for robo-advising. 
Building on the basic features of the model studied in \citet{capponi2022personalized}, we study the interaction between a robo-advisor that is in charge of trading and at all times informed about the market parameters and a human client with stochastic, time-varying preferences. 

Specifically, we consider a binomial tree model for the financial market as outlined in Section \ref{sec:Model} and suppose that the robo-advisor has accurate knowledge of $\mathcal{F}_{i-1}$-measurable market parameters $(u_i,d_i,p_i)$ for each trading period $[i-1, i)$.
The preferences of the human client are described by a random utility process $U^C_i (x, \omega), i\in\mathbb{N}_0$, but the robo-advisor only knows these preferences at times when there is an interaction with the client. 

In an ideal world, the robo-advisor would know the preferences of the agent at each point in time. In this case, at every time $i$, the robo-advisor solves a $1$-period forward problem based on the initial utility $U^C_i$ and current market $(u_i,d_i,p_i)$ resulting in a strategy $\pi^I_i(x)$ for given wealth $X_{i-1}=x$. 
However, in the real world, the client will not spend too much time and energy interacting with the robo-advisor, and there are thus times where the robo-advisor does not know the current preferences of the client.

At time zero, after setting up an account, the client communicates  her initial preferences and her interaction schedule to the robo-advisor and agrees on the assessment of market parameters provided by robo-advisor. 
We suppose that the robo-advisor, while not knowing the current preferences of the client, acts in the best interest of the client and updates preferences according to a forward performance process.
This assures that investment is consistent with the initial preferences and the last assessment of the financial market approved by the client. 
The robo-advisor thus determines the forward process according to the scheme developed in Section \ref{sec:FunctionalEquation} and invests period-by-period until the next interaction with the client.

We assume that client and robo-advisor only interact at times $km$, $k \in \mathbb{N}_0$. 
At each interaction time, the robo-advisor takes $U^C_{km}$ and current market parameters $(u_{km}, d_{km}, p_{km})$ as inputs and determines the  $m$-forward $U^{mF}_{(k+1)m}$ computed with time-homogeneous parameters together with the strategy profile $\pi^{mF}_t$ for $t = k m + 1, \dots, (k+1) m$. 

In our setting, there thus results in an ideal strategy profile $\pi^I$ and the actual strategy profile $\pi^{mF}$ implemented by the robo-advisor.
Our goal is to control the diversion between the two by choosing an optimal interaction schedule.
To this end, we define the first time where the scaled absolute deviation between $\pi^I$ and $\pi^{mF}$ exceeds a certain value
\begin{align}\label{MeasureRobo1}
    m_{\kappa} = \min \left\{ m \in \mathbb{N} \bigg\vert \max_{\ell \in \{1, \dots , m\}} \sup_{x \geq 0} \left\vert \frac{\pi^I_\ell (x) - \pi^{m F}_\ell (x)}{\sigma(x)} \right\vert > \kappa
    \right\},
\end{align}
where $\kappa$ is a tolerance parameter and $\sigma (x)$ is a scaling function.
Note that $m_{\kappa}$ is a random variable that depends on how the actual market parameters $(u_t,d_t,p_t)$ and client preferences $U^C$ evolve. 

In this paper, we will focus on the updating of the probability $p$ of an upward movement of the stock and risk Arrow-Pratt measure of risk aversion $1/ \theta$ of a client with preferences belonging to the family of CRRA utility functions. 
We assume that the possible values of the return of the stock, $u$ and $d$, are constant and identical for both ideal and realistic situations.
Working within the family of CRRA utility functions allows us to directly apply Theorem \ref{thm:mForwardConstructionHeterogeneous} unless $\theta = - \log_{a} b$. 
However, even in this special case, only uniqueness is lost and we can still work with the canonical choice for the $m$-forward belonging to the same family of CRRA utility functions.  


\subsection{Optimal interaction schedule under robust approach}
In the following, we study a robust approach of determining the optimal interaction schedule $m^*$. 
Similarly as in the robust approach to determine an optimal interaction schedule between an automated trading system and human expert studied in Section \ref{subsec:ATS_Robust}, we allow for any predictable processes $(p_i)_{i \in \mathbb{N}}$ and $(\theta_i)_{i \in \mathbb{N}}$ that remain within reasonable intervals specified at the beginning of each evaluation period, and then compute the difference with the ideal strategy corresponding to the worst possible specification this distribution could take.

The transition probability and the risk preference parameter for each trading period are given according to the initial values $p$ and $\theta$, by $p_i=D_ip$ and $\theta_i=E_i\theta$, for $\mathcal{F}_{i-1}$-measurable random variables $D_i$ and $E_i$ which can take value in the intervals $[D_{i,d}, D_{i,u}]$ and $[E_{i,d}, E_{i,u}]$ respectively.
The upper and lower bounds 
$D_{i,d}$, $D_{i,u}$, $E_{i,d}$ and $E_{i,u}$ are assumed to be periodic in the evaluation period and reflect that uncertainty about model and preference parameters increase as the time since the last point of contact increases. 
Formally, we suppose that
$D_{i,d} = \tilde{D}_{({i \mod m}),d}$, $D_{i,u}= \tilde{D}_{({i \mod m}), u}$, $E_{i,d} = \tilde{E}_{({i \mod m}),d}$, and $E_{i,u}= \tilde{E}_{({i \mod m}), u}$ for $i \in \mathbb{N}$, for exogenously given sequences $(\tilde{D}_{i,d})_{i\in \mathbb{N}}$, $(\tilde{D}_{i,u})_{i \in \mathbb{N}}$, $(\tilde{E}_{i,d})_{i \in \mathbb{N}}$ and $(\tilde{E}_{i,u})_{i \in \mathbb{N}}$  satisfying that  $\tilde{D}_{i,u}p < 1$, $\tilde{E}_{i,u}\theta \neq 1$, $\tilde{E}_{i,d}\theta \neq 1$, and $\tilde{D}_{i,u}$ and $\tilde{E}_{i,u}$ are increasing while $\tilde{D}_{i,d}$ and $\tilde{E}_{i,d}$ are decreasing in $i \in \mathbb{N}$.
We also assume that $\tilde D_{1,u}=\tilde D_{1,d}= \tilde E_{1,u}=\tilde E_{1,d} = 1$, i.e., the robo-advisor operates based on accurate information about the market and the preferences of the client whenever there is an interaction.

Considering that for each trading period $[i-1, i)$ the absolute deviation between actual and ideal investment policy, $\left\vert \pi^I_i (x) - \pi^{m F}_i (x) \right\vert$, is an $\mathcal{F}_{i-1}$-measurable random variable,
we next characterize how the optimal interaction schedule is defined according to a robust criterion.
\begin{definition}
The optimal interaction schedule under the robust approach is defined by 
\begin{align}\label{RobustMeasure}
    m^{*}_{\kappa} = \essinf_{\omega \in \Omega}  m_\kappa =  \min \left\{ m \in \mathbb{N} \bigg\vert \max_{\ell \in \{1, \dots , m\}} \esssup_{\omega \in \Omega} \sup_{x \geq 0} \left\vert \frac{\pi^I_\ell (x) - \pi^{m F}_\ell (x)}{\sigma(x)} \right\vert > \kappa
    \right\},
\end{align}
where $m_\kappa$ is defined by $(\ref{MeasureRobo1})$.
\end{definition}
Determining an optimal interaction schedule according to the robust approach ensures that the largest possible value of absolute deviation between the ideal and implemented strategy after scaling remains within some acceptable level that is pre-specified by the client. 

In our analysis, price levels $u$ and $d$ of the return $R_i, i\in \mathbb{N}_0$ remain unchanged and are estimated at the beginning of the investment process. 
Let $a_i=\frac{1-p_i}{p_i}\frac{q}{1-q}, b=\frac{1-q}{q}, c_i=\frac{1-p_i}{1-q}$, and $\delta = \frac{1+b}{c_1^\theta(a_1^{-\theta}+b)}$. 
The following proposition formulates the maximum possible deviation between ideal and implemented strategy. 

\begin{proposition}\label{prop:BoundDeviationSchedule}
Suppose that the initial datum is of the form $U_0(x)=(1-\frac{1}{\theta})^{-1}x^{1-\frac{1}{\theta}},$ $x>0$, for some $1 \neq \theta>0$, and let $(D_i)_{i\in \mathbb{N}}$ and $(E_i)_{i\in \mathbb{N}}$ be predictable processes taking values in $[D_{i,d},D_{i,u}]$ and $[E_{i,d},E_{i,u}]$, where $D_{i,d}, D_{i,u}, E_{i,d}$ and $E_{i,u}$ satisfy the assumption of periodicity in the interaction schedule $m$ and are such that absence of arbitrage is maintained.
We then have that
\begin{align}\label{MaxDeviation}
\begin{split} 
\esssup_{\omega \in \Omega } \left\vert \pi^I_i (x) - \pi^{m F}_i (x) \right\vert
&  = 
\frac{x}{u-d} \max\big\{{\boldsymbol G}(D_{i,u}p,E_{i,u}\theta\mathbbm{1}_{\{D_{i,u}p>q\}}+E_{i,d}\theta\mathbbm{1}_{\{D_{i,u}p<q\}}) - \boldsymbol G_0, \\
& \hspace{3cm} \boldsymbol G_0- \boldsymbol G(D_{i,d}p,E_{i,d}\theta\mathbbm{1}_{\{D_{i,d}p>q\}}+E_{i,u}\theta\mathbbm{1}_{\{D_{i,d}p<q\}}) \big\},
\end{split}
\end{align}
where $\boldsymbol G(\tilde{p},\tilde{\theta}):=\frac{q^{-\tilde{\theta}}\tilde{p}^{\tilde{\theta}}-(1-q)^{-\tilde{\theta}}(1-\tilde{p})^{\tilde{\theta}}}{q^{1-\tilde{\theta}}\tilde{p}^{\tilde{\theta}}+(1-q)^{1-\tilde{\theta}}(1-\tilde{p})^{\tilde{\theta}}}$ and $\boldsymbol G_0=\frac{q^{-\theta}p^{\theta}-(1-q)^{-\theta}(1-p)^{\theta}}{q^{1-\theta}p^{\theta}+(1-q)^{1-\theta}(1-p)^{\theta}}$. Furthermore, $\boldsymbol G(\tilde{p},\tilde{\theta})$ increases in $\tilde{p}$ and increases in $\tilde{\theta}$ when $\tilde{p}>q$ respectively decreases in $\tilde{\theta}$ when $\tilde{p}<q$, and $\esssup_{\omega \in \Omega }\left\vert\pi^I_i (x) - \pi^{m F}_i (x)\right\vert$ is increasing in $i \in \mathbb{N}$.
\end{proposition}

Noticing that both strategies $\pi^I$ and $\pi^{mF}$ are proportional in wealth motivates to consider the scaling function $\sigma(x) = x$.
This then removes the dependence on wealth and thus leads to more tractable results. 
We also know from Proposition \ref{prop:BoundDeviationSchedule} that the maximum absolute deviation is increasing in time, and, when $\sigma(x) = x$, $m^*_{\kappa}$ defined in (\ref{RobustMeasure}) can thus be reduced to 
\begin{align}\label{OptimalScheduleRobo}
    m^{*}_{\kappa} = \min\left\{ m \in \mathbb{N} \big\vert \text{either condition } (C_{m,\kappa}^1) \text{ or } (C_{m,\kappa}^2) \text{ is satisfied}
    \right\},
\end{align}
where condition $(C_{m,\kappa}^1)$ refers to 
\begin{align*}
    {\boldsymbol G}(D_{m,u}p,E_{m,u}\theta\mathbbm{1}_{\{D_{m,u}p>q\}}+E_{m,d}\theta\mathbbm{1}_{\{D_{m,u}p<q\}})> \boldsymbol G_0+\kappa(u-d),
\end{align*}
while condition $(C_{m,\kappa}^2)$ refers to 
\begin{align*}
    \boldsymbol G(D_{m,d}p,E_{m,d}\theta\mathbbm{1}_{\{D_{m,d}p>q\}}+E_{m,u}\theta\mathbbm{1}_{\{D_{m,d}p<q\}})< \boldsymbol G_0-\kappa(u-d) .
\end{align*}
Condition $(C_{m,\kappa}^1)$ corresponds to the case where the upper range of uncertainty about probabilities of upward moves in the stock triggers interaction, whereas interaction is triggered by the lower range of the uncertainty range if $(C_{m,\kappa}^2)$ holds first. 

The measure $m^{*}_{\kappa}$  depends on 
the market parameters $p$, $u$, and $d$, the Arrow-Pratt measure of relative risk aversion $1/\theta$ of the initial datum of an agent, the tolerance level $\kappa$, and the uncertainty about the evolution of future beliefs captured in the sequences $(D_{i,d}, D_{i,u})$ and $(E_{i,d}, E_{i,u})$. 
In practice, at time zero, we choose an optimal interaction schedule based on our current understanding of the market. 
At the subsequent interaction time, we update the market parameters, and then choose a new optimal interaction schedule. 
Therefore, market parameters, updating frequencies, preferences and investment strategies move together forward in time.

By the monotonicity of the function $\boldsymbol G$, ${\boldsymbol G}(D_{m,u}p,E_{m,u}\theta\mathbbm{1}_{\{D_{m,u}p>q\}}+E_{m,d}\theta\mathbbm{1}_{\{D_{m,u}p<q\}})$ is increasing while $\boldsymbol G(D_{m,d}p,E_{m,d}\theta\mathbbm{1}_{\{D_{m,d}p>q\}}+E_{m,u}\theta\mathbbm{1}_{\{D_{m,d}p<q\}})$ is decreasing as the length of the evaluation period $m$ increases.
This is due to the increase in the uncertainty about parameters  since the last interaction time captured in the increasing upper bounds $D_{m,u}, E_{m,u}$ and the decreasing lower bounds $D_{m,u}$ and $E_{m,u}$. 
Therefore, the optimal interaction schedule under the robust approach is finite if and only if
\begin{align*}
\begin{split} &  
\kappa< \lim \limits_{T\rightarrow +\infty}\frac{1}{u-d} \max\big\{{\boldsymbol G}(\tilde{D}_{T,u}p,\tilde{E}_{T,u}\theta\mathbbm{1}_{\{\tilde{D}_{T,u}p>q\}}+\tilde{E}_{T,d}\theta\mathbbm{1}_{\{\tilde{D}_{T,u}p<q\}}) - \boldsymbol G_0, \\
& \hspace{3cm} \boldsymbol G_0- \boldsymbol G(\tilde{D}_{T,d}p,\tilde{E}_{T,d}\theta\mathbbm{1}_{\{\tilde{D}_{T,d}p>q\}}+\tilde{E}_{T,u}\theta\mathbbm{1}_{\{\tilde{D}_{T,d}p<q\}}) \big\} .
\end{split}
\end{align*}


Next, we investigate how the optimal interaction schedule $m^*$ depends on the tolerance parameter $\kappa$ and the uncertainty about the transition probability $p$ and risk aversion $\theta$.
First, it follows immediately from the definition that $m_{\kappa}^*$ is increasing in $\kappa$.
This is consistent with our intuition that when the client can tolerate with the inaccuracy of investment strategy to a larger extent, robo-advisors would reduce the frequency of interaction accordingly. 

When there is higher uncertainty about the future model or preference parameters $p$ respectively $\theta$, i.e., when $D_{i,u}$ or $E_{i,u}$ increase and/or  $D_{i,d}$ or $E_{i,d}$ decrease, ${\boldsymbol G}(D_{i,u}p,E_{i,u}\theta\mathbbm{1}_{\{D_{i,u}p>q\}}+E_{i,d}\theta\mathbbm{1}_{\{D_{i,u}p<q\}})$ increases and $\boldsymbol G(D_{i,d}p,E_{i,d}\theta\mathbbm{1}_{\{D_{i,d}p>q\}}+E_{i,u}\theta\mathbbm{1}_{\{D_{i,d}p<q\}})$ decreases by the monotonicity of the function $\boldsymbol G$ derived in Proposition \ref{prop:BoundDeviationSchedule}. 
Hence, the optimal interaction schedule $m^*_{\kappa}$ decreases implying that if there is greater uncertainty about model or preference parameters, one should interact more frequently. 
In contrast to the previous application on automated trading where a uniform increase was required, an increase in uncertainty generally leads to more frequent optimal interaction in this application. 

\begin{figure}[H]
   \includegraphics[width=7.96cm]{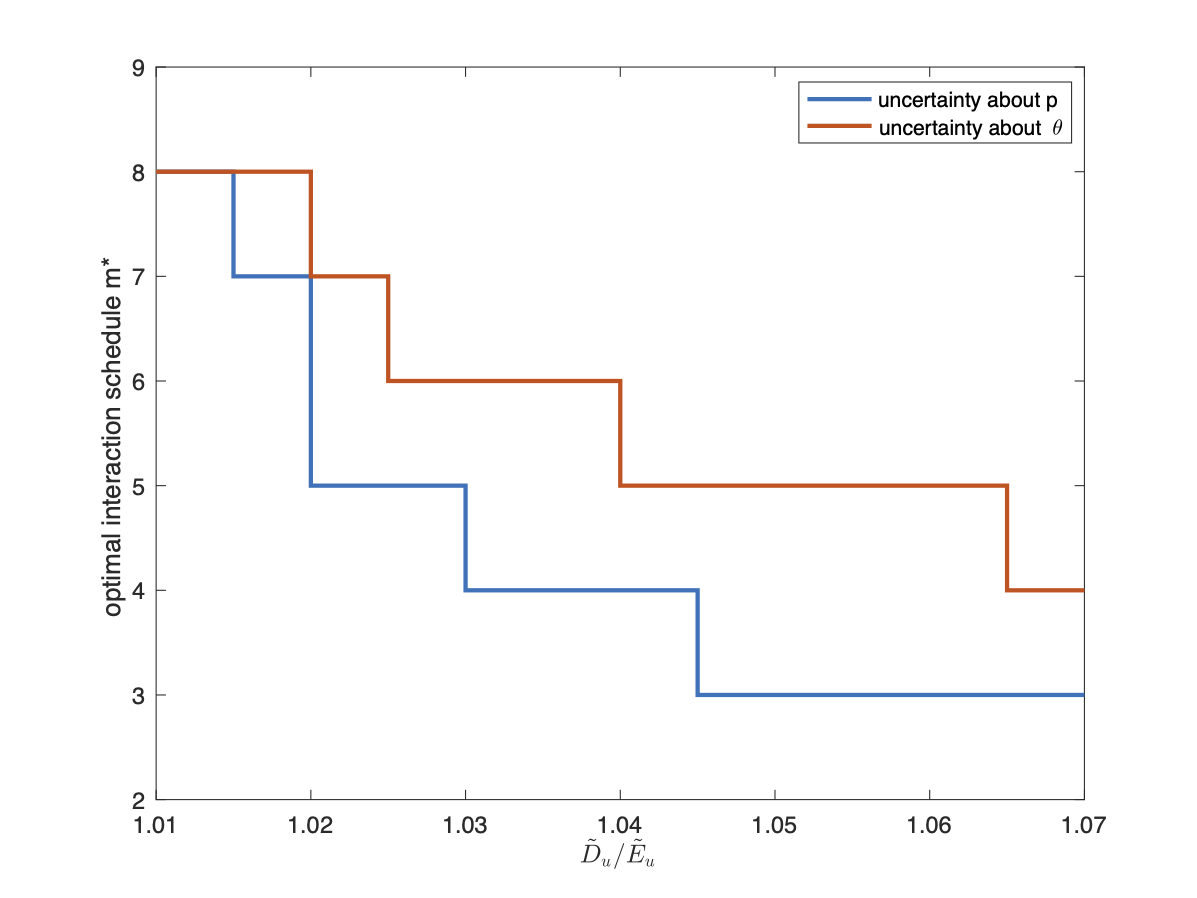}
  \includegraphics[width=7.96cm]{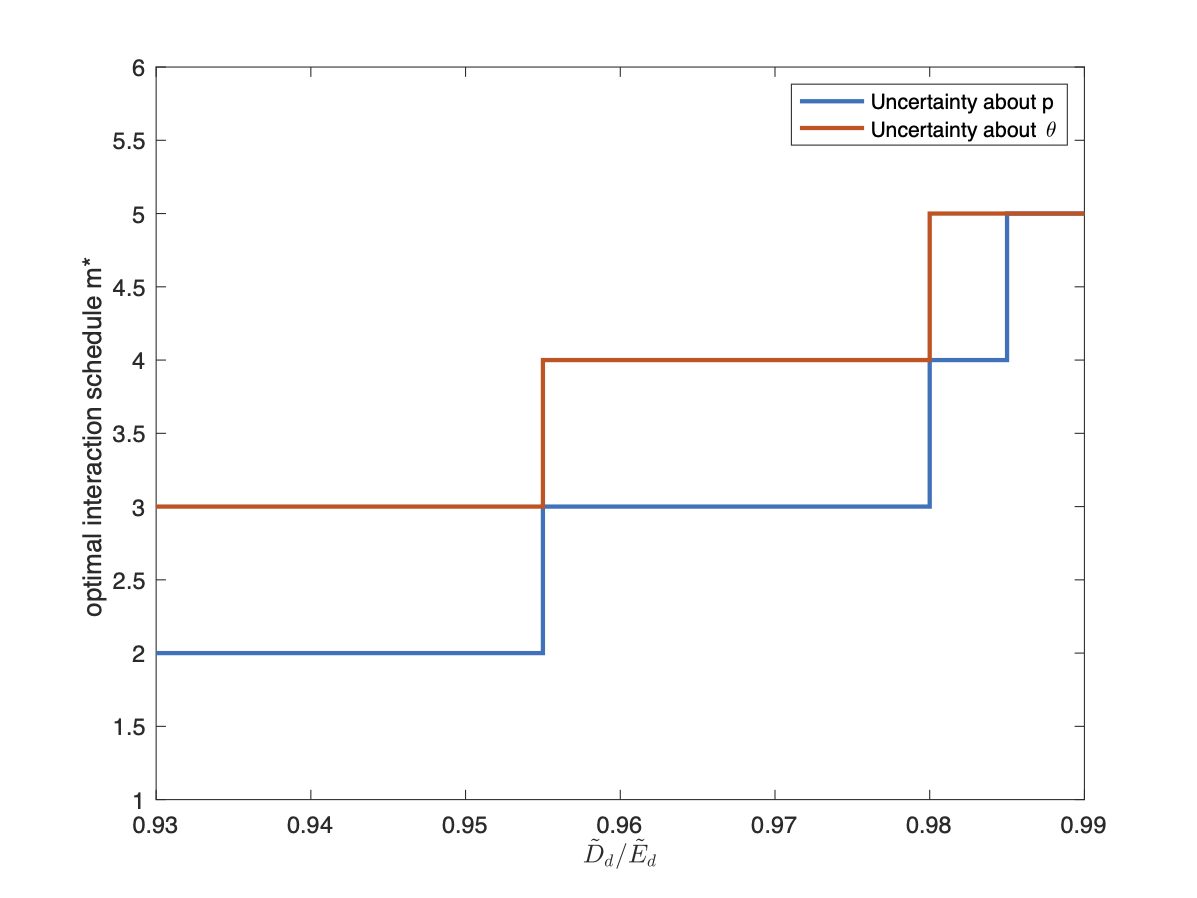}
  \caption{Impact of increasing uncertainty about $p$ or $\theta$.}
  \label{figUncert}
\end{figure}

We next investigate whether the optimal interaction schedule $m^{*}$ is more sensitive with an increase in uncertainty about an upward move for the stock or an increase in uncertainty about the risk preferences of the client. 
Suppose that the market parameters are $p=0.6$, $u=1.15$, $d=0.9$, the risk aversion parameter of the initial datum is $\theta=5$, the client tolerance is $\kappa=0.3$, and the bounds for the updating of future probabilities and risk aversion are given by $D_{i,d}=\widetilde{D}_d^{\mod(i-1,m)}, D_{i,u}= \widetilde{D}_u^{\mod(i-1,m)}, E_{i,d}= \widetilde{E}_d^{\mod(i-1,m)}, E_{i,u} = \widetilde{E}_u^{\mod(i-1,m)}$, where $\widetilde{D}_d=\widetilde{E}_d=1$.
On the left-hand side of Figure \ref{figUncert}, we increase only $\widetilde{D}_u$ and fix $\widetilde{E}_u=1.01$ for studying the effect of increasing uncertainty about $p_i$ (the blue line), while we increase only $\widetilde{E}_u$ and fix $\widetilde{D}_u=1.01$ for studying the sensitivity with respect to the uncertainty about risk aversion $\theta_i$ (the red line). 
Analogously, on the right-hand side of Figure \ref{figUncert}, we set $\widetilde{D}_u=\widetilde{E}_u=1$, and vary only $\widetilde{D}_d$ with $\widetilde{E}_d=0.99$ fixed or only $\widetilde{E}_d$ with $\widetilde{D}_d=0.99$ fixed.
From the numerical analysis shown in Figure \ref{figUncert}, we infer that, while an increase in the uncertainty about $p$ or $\theta$ leads to more frequent optimal interaction, the optimal interaction schedule is more sensitive to uncertainty in the market parameter than about the risk preferences of the agent.

We further find that it is optimal to interact less frequently in a more volatile environment with all else being equal. 
We increase $u$ and adjust $d=\frac{1-qu}{1-q}$ accordingly such that the risk-neutral probability $q$ remains constant but the stock becomes more volatile. This does not alter the values of ${\boldsymbol G}(D_{m,u}p,E_{m,u}\theta\mathbbm{1}_{\{D_{m,u}p>q\}}+E_{m,d}\theta\mathbbm{1}_{\{D_{m,u}p<q\}})$, $\boldsymbol G(D_{m,d}p,E_{m,d}\theta\mathbbm{1}_{\{D_{m,d}p>q\}}+E_{m,u}\theta\mathbbm{1}_{\{D_{m,d}p<q\}})$, and ${\boldsymbol G}_0$, but $u-d$ increases. 
One can thus easily conclude from $(\ref{OptimalScheduleRobo})$ that $m_{\kappa}^*$
increases as conditions $(C_{m,\kappa}^1)$ and $(C_{m,\kappa}^2)$ can only be satisfied at later times. 
The intuition for this is that the scale of risky investment decreases in a more volatile market, and the difference between the actual and ideal strategy thus also decreases.

\section{Conclusions}\label{sec:Conclusions}
We studied discrete-time predictable forward processes when trading dates do not coincide with performance evaluation dates in a binomial tree model for the financial market. 
Our main technical contributions are conditions for existence and uniqueness as well as explicit solutions for the functional equations associated with the construction of predictable forward processes. 
We then applied the results to study the asset allocation problem faced by automated trading platforms and robo-advisors, the applications where performance evaluation naturally occurs at a lower frequency than trading. 
Our findings and discussions show that predictable forward performance processes constitute a viable framework to model preferences of agents of automated trading and robo-advisors and can provide valuable insights when determining an optimal interaction schedule between the machine and its human clients.  


 \section*{Acknowledgments}
The authors are grateful to Bahman Angoshtari, John Armstrong, Agostino Capponi, Eric C.K. Cheung, Tahir Choulli, Martin Herdegen, Pedro Valls Pereira, Yang Shen, Thaleia Zariphopoulou, an anonymous associate editor, and two anonymous referees for their helpful comments and suggestions.
This work was presented at the Annual Meeting of the Society of Management Science and Engineering of China, Workshop on Advances in Optimal Decision Making under Uncertainty at the University of Chicago, The 2nd London Oxford Warwick Financial Mathematics Workshop, Bachelier World Congress 2022, Quantitative Finance and Financial Econometrics 2022, 2022 SIAM Annual Meeting, Kickoff Conference of the ANR Project DREAMeS, Professor Duan Li Memorial Workshop on Finance, Optimization and Control at Sun Yat-Sen University as well as at a seminar presentation at the University of New South Wales. 
The authors thank the participants for their comments.
Moris Strub gratefully acknowledges funding through the National Natural Science Foundation of China under Grant No. 72050410356.
Author order is alphabetical. All authors are co-first authors of this paper. 

\begin{appendices}

\section*{Appendix. Proofs}
\renewcommand{\thesubsection}{\Alph{subsection}}

\subsection{Proof of Theorem \ref{thm:mForwardConstructionHeterogeneous}}


As discussed in Section 2, if two utility functions $U_0$ and $U_m$ solve problem \eqref{eq:InverseInvestmentProblem}, then their associated inverse marginals satisfy \eqref{eq:FctEQ_Multinomial_Heterogeneous}. Conversely, when a pair of inverse marginal functions $I_0$ and $I_m$ satisfy \eqref{eq:FctEQ_Multinomial_Heterogeneous}, then the corresponding utility functions satisfy \eqref{eq:InverseInvestmentProblem} up to a constant. Theorem 2.4 in \cite{strub2021evolution} together with the subsequent discussion therein shows that $U_m$ defined as in Theorem \ref{thm:mForwardConstructionHeterogeneous} does indeed solve \eqref{eq:InverseInvestmentProblem} when $I_m$ solves \eqref{eq:FctEQ_Multinomial_Heterogeneous}.
Moreover, the expression for the optimal wealth $X_m^*$ follows from the existing theory on classical expected utility maximization once we obtained $U_m$ and regard \eqref{eq:InverseInvestmentProblem} as a classical, backward problem. 
Therefore, it remains to show that, under the assumption that $\{(\alpha_1,\dots,\alpha_i ) \}_{i=1,\dots, m}$ exists, $I_m$ given in \eqref{eq:ForwardInverseMarginal_Heterogeneous}  is the unique solution to \eqref{eq:FctEQ_Multinomial_Heterogeneous} in the class of inverse marginal functions.

Making the substitution $\hat{y} =\prod \limits_{i=1}^m\frac{(1-p_i)}{(1-q_i)}y$  allows us to transform \eqref{eq:FctEQ_Multinomial_Heterogeneous} to
\begin{align*}
I_0 \left( \prod \limits_{i=1}^m\frac{(1-p_i)}{(1-q_i)}y \right) = \sum\nolimits_{j=0}^{2^m-1}\prod \limits_{i=1}^mq_i^{\gamma_{j,i}}(1-q_i)^{1-{\gamma_{j,i}}}I_m \left(\prod \limits_{i=1}^m\frac{q_i^{\gamma_{j,i}}(1-p_i)^{\gamma_{j,i}}}{p_i^{\gamma_{j,i}}(1-q_i)^{\gamma_{j,i}}}y\right).
\end{align*}
Next, we multiply both sides by $\left(\prod \limits_{i=1}^mq_i \right)^{-1}$ and recall the expression in terms of $a_i,b_i,c_i, i=1,2,...,m$ to obtain
\begin{align}
    \prod \limits_{i=1}^m(1+b_i)I_0\left(\prod \limits_{n=1}^mc_ny\right)=\sum\nolimits_{j=0}^{2^m-1}\prod \limits_{i=1}^mb_i^{1-\gamma_{j,i}}I_m\left(\prod \limits_{k=1}^ma_k^{\gamma_{j,k}}y\right).\label{16}
\end{align}

We notice that when market parameters are time-heterogeneous, the arguments of $I_m$ are not in the form of iterate functions. 
We therefore cannot use standard techniques for integral equations.
Instead, we aim to show by mathematical induction that if, for a given $I_0$, there exist functions $(\tilde I_i)_{i=1}^{m}$ such that they satisfy a system of equations $\tilde I_i(a_iy)+b_i\tilde I_i(y)=(1+b_i)\tilde I_{i-1}(c_iy)$, $i=1,2,\dots,m$,  then $I_0$ and $I_m=\tilde I_m$ satisfy $(\ref{16})$.


First, when $m=1$, the statement naturally holds. Let us then assume that the statement is true for $m=M$. When $m=M+1$, the left hand side of equation $(\ref{16})$ becomes
\begin{align}\label{proveequivalence}
\begin{split}
    &\prod \limits_{i=1}^{M+1}(1+b_i)I_0\left(\prod \limits_{n=1}^{M+1}c_ny\right)\\
    =&(1+b_{M+1})\prod \limits_{i=1}^{M}(1+b_i)I_0\left(\prod \limits_{n=1}^Mc_nc_{M+1}y\right)\\
    =&(1+b_{M+1})\sum\nolimits_{j=0}^{2^M-1}\prod \limits_{i=1}^Mb_i^{1-\gamma_{j,i}}\tilde I_M\left(\prod \limits_{k=1}^Ma_k^{\gamma_{j,k}}c_{M+1}y\right)\\
    =&\sum\nolimits_{j=0}^{2^M-1}\prod \limits_{i=1}^Mb_i^{1-\gamma_{j,i}}\left(\tilde I_{M+1} \left(a_{M+1}\prod \limits_{k=1}^Ma_k^{\gamma_{j,k}}y\right) + b_{M+1}\tilde I_{M+1} \left(\prod \limits_{k=1}^Ma_k^{\gamma_{j,k}}y\right)\right).
\end{split}
\end{align}
Note that the right hand side of (\ref{16}) is given by $\sum\nolimits_{j=0}^{2^{M+1}-1}\prod \limits_{i=1}^{M+1}b_i^{1-\gamma_{j,i}}I_{M+1}\left(\prod \limits_{k=1}^{M+1}a_k^{\gamma_{j,k}}y\right)$ when setting $m=M+1$. 
Observe that for $j\in \{0,1,...,2^M-1\}$, we  have $\gamma_{j,M+1}=0$, where $\gamma_{j,M+1}$ is the first digit of $j$ in binary form if expressed in $M+1$ digits in total. 
Thus, the term inside the summation becomes $b_{M+1}\prod \limits_{i=1}^Mb_i^{1-\gamma_{j,i}}I_{M+1}\left(\prod \limits_{k=1}^Ma_k^{\gamma_{j,k}}y\right)$.
However, for $j\in \{2^M,2^M+1,...,2^{M+1}-1\}$, $\gamma_{j,M+1}=0$, and the term inside the summation is given by $\prod \limits_{i=1}^Mb_i^{1-\gamma_{j,i}}I_{M+1}\left(a_{M+1}\prod \limits_{k=1}^Ma_k^{\gamma_{j,k}}y\right)$. 
Hence, the right hand side of (\ref{16}) is equal to the last line of (\ref{proveequivalence}) by letting $I_m=\tilde I_m$. 
This proves the claim for $m=M+1$, and thus for arbitrary $m$ by induction.

For the other direction, note that with the substitution $\tilde I_{k-1}(y)=\frac{1}{1+b_k} \left(\tilde I_k(\frac{a_ky}{c_k})+b_k\tilde I_k(\frac{y}{c_k}) \right)$, $k=m,\dots,1$, equation 
\begin{align*}
    \prod \limits_{i=1}^k(1+b_i)I_0\left(\prod \limits_{n=1}^kc_ny\right)=\sum\nolimits_{j=0}^{2^k-1}\prod \limits_{i=1}^kb_i^{1-\gamma_{j,i}}\tilde I_k\left(\prod \limits_{n=1}^ka_n^{\gamma_{j,n}}y\right)
\end{align*} 
becomes
\begin{align*}
    \prod \limits_{i=1}^{k-1}(1+b_i)I_0\left(\prod \limits_{n=1}^{k-1}c_ny\right)=\sum\nolimits_{j=0}^{2^{k-1}-1}\prod \limits_{i=1}^{k-1}b_i^{1-\gamma_{j,i}}\tilde I_{k-1}\left(\prod \limits_{n=1}^{k-1}a_n^{\gamma_{j,n}}y\right) .
\end{align*}
 Hence, we can show that if $I_0$ and $I_m$ are related by $(\ref{16})$, then there exist functions $(\tilde I_i)_{i=1}^{m-1}$ such that they satisfy a system of equations $\tilde I_i(a_iy)+b_i\tilde I_i(y)=(1+b_i)\tilde I_{i-1}(c_iy), i=1,2,...,m$ with $\tilde I_0=I_0$ and $\tilde I_m=I_m$.

Solving $\tilde I_i(a_iy)+b_i\tilde I_i(y)=(1+b_i)\tilde I_{i-1}(c_iy), i=1,2,...,m$ by sequentially and repeatedly applying Theorem 6.3 in \cite{angoshtari2020predictable} adapted to the notation herein, we can finally derive the expression of $I_m$
and obtain the claimed conditions for uniqueness. 

We can show by induction that $\tilde I_i$, $i=1,2,\dots,m$ 
solving $\tilde I_i(a_iy)+b_i\tilde I_i(y)=(1+b_i)\tilde I_{i-1}(c_iy)$, $i=1,2,\dots,m$,
must be given by
\begin{align}\label{eq:TildeI_i_Heterogeneous}
    \tilde I_i(y) = \frac{\prod \limits_{v=1}^i(1+b_v)}{\prod \limits_{j=1}^ib_j^{\alpha_j}} \sum\limits_{n_1=0,...,n_i=0}^{\infty}(-1)^{p_{(n_1,\dots,n_i)}} \prod\limits_{k=1}^ib_k^{n_k(1-2\alpha_k)} I_0 \left( \prod\limits_{s=1}^ia_s^{n_s(2\alpha_s-1)+(\alpha_s-1)}\prod\limits_{u=1}^ic_uy \right).
\end{align} 
We outline these arguments below.
First, the statement obviously holds for $i=1$ by Theorem 6.3 in \cite{angoshtari2020predictable}.
In the general inductive step, we show that if $\tilde I_k$ is given by \eqref{eq:TildeI_i_Heterogeneous} with its corresponding $(\alpha_1, \dots, \alpha_k)$, then the statement is also true for $\tilde I_{k+1}$. Noticing that $\Phi_{k}^{(\alpha_1, \dots, \alpha_k)}$ and $\Psi_{k}^{(\alpha_1, \dots, \alpha_k)}$ defined for solving the single-period inverse problem with $\tilde I_k$ given by \eqref{eq:TildeI_i_Heterogeneous} must be expressed by \eqref{eq:DefPhiPsi_heter}, we apply \cite[Theorem 6.3]{angoshtari2020predictable} to obtain that $\tilde I_{k+1}$ is unqiue and given by
\begin{align*}
    \tilde I_{k+1}(y)=\frac{\prod \limits_{v=1}^{k+1}(1+b_v)}{b_{k+1}\prod \limits_{j=1}^kb_j^{\alpha_j}}\sum\limits_{n_1=0,...,n_{k+1}=0}^{\infty}&\bigg((-1)^{p_{(n_1,\dots,n_{k+1})}}b_{k+1}^{-n_{k+1}}\prod\limits_{t=1}^kb_t^{n_t(1-2\alpha_t)}\\
    & \qquad \times I_0 \left( a_{k+1}^{n_{k+1}}\prod\limits_{s=1}^ka_s^{n_s(2\alpha_s-1)+(\alpha_s-1)}\prod\limits_{u=1}^{k+1}c_uy \right) \bigg), 
\end{align*} 
if $\left( \Phi_{k}^{(\alpha_1, \dots, \alpha_k)}, \Psi_{k}^{(\alpha_1, \dots, \alpha_k)} \right)$ satisfies $(C1)$, and given by
\begin{align*}
    \tilde I_{k+1}(y)=\frac{\prod \limits_{v=1}^{k+1}(1+b_v)}{\prod \limits_{j=1}^kb_j^{\alpha_j}}\sum\limits_{n_1=0,...,n_{k+1}=0}^{\infty}&\bigg((-1)^{p_{(n_1,\dots,n_{k+1})}}b_{k+1}^{n_{k+1}}\prod\limits_{t=1}^kb_t^{n_t(1-2\alpha_t)}\\
    & \qquad \times I_0(a_{k+1}^{-(n_{k+1}+1)}\prod\limits_{s=1}^ka_s^{n_s(2\alpha_s-1)+(\alpha_s-1)}\prod\limits_{u=1}^{k+1}c_uy)\bigg), 
\end{align*} 
if $\left( \Phi_{k}^{(\alpha_1, \dots, \alpha_k)}, \Psi_{k}^{(\alpha_1, \dots, \alpha_k)} \right)$ satisfies $(C2)$.
Therefore, $\tilde I_{k+1}(y)$ can be expressed by \eqref{eq:TildeI_i_Heterogeneous} with the sequence given by $(\alpha_1, \dots, \alpha_k,1)$ when $(C1)$ is satisfied or $(\alpha_1, \dots, \alpha_k,0)$ when $(C2)$ is satisfied. This proves the claim and thus shows that $I_m$ is uniquely given by \eqref{eq:ForwardInverseMarginal_Heterogeneous}.
\qed 

\subsection{Proof of Corollary \ref{cor:Measurability_Heterogeneous}}
Let $x > 0$.
We first note that vectors $(a_i)_{i=1}^m,(b_i)_{i=1}^m,$ and $(c_i)_{i=1}^m$ are all Borel-measurable functions of the market parameters $(p,u,d) \in \mathcal{M}$. 
We will drop the classifier Borel- for the remainder of this proof.
First, we prove the measurability of every $\alpha_i,i=1,\dots,m$  by simple induction.
 
The base case, $i=1$, is trivial since $\alpha_1$ is constant.
We then assume that $(\alpha_k)_{k=1}^{i}$ is measurable, and prove measurability of $\alpha_{i+1}$. By its definition and the assumption that $\{(\alpha_1,\dots,\alpha_i)\}_{i=1,\dots,m}$ exists, $\alpha_{i+1}$ can be expressed as $\alpha_{i+1}=\mathbbm{1}_{\{\left( \Phi_{i}^{(\alpha_1,\dots,\alpha_i)}, \Psi_{i}^{(\alpha_1,\dots,\alpha_i)} \right) \enskip \mathrm{satisfies} \enskip (C1)\}}$
By Theorem 6.3 in \cite{angoshtari2020predictable} and since $I_0$ is continuously differentiable, the infinite series  of $(\Phi_i^{(\alpha_1,\dots,\alpha_i)})^{'}$ and $\Psi_i^{(\alpha_1,\dots,\alpha_i)}$ converge for $(p,u,d)\in\mathcal{M}$.
Therefore, $(\Phi_i^{(\alpha_1,\dots,\alpha_i)})^{'}$ and $\Psi_i^{(\alpha_1,\dots,\alpha_i)}$ defined in (\ref{eq:DefPhiPsi_heter}) are measurable as pointwise limits of measurable functions, which in turn shows that $\alpha_{i+1}$ is measurable.

Note that the series of $I_m$ is derived by sequential application of Theorem 6.3 in \cite{angoshtari2020predictable}. Since all intermediate functions are shown to be convergent, so does $I_m$. 
The measurable dependence of $I_m$ on the market parameters then follows from the explicit expression in \eqref{eq:ForwardInverseMarginal_Heterogeneous} as a pointwise limit of measurable function in a converging series. 

In a multi-period binomial market, the expectation in the expression of $U_m$ is essentially a finite sum of integral terms. Given that the inverse function of a strictly monotone function is also strictly monotone and thus integrable over finite intervals, its corresponding integral with variable lower limit of integration $f(u)=\int_{u}^{x} I_m^{-1}(t)dt$ for any given $x$ exists and is continuous. Therefore, the integral terms $f(I_m(\frac{d\mathbb{Q}}{d\mathbb{P}}U_0'(1)))$, are compositions of two measurable functions, which then shows that $U_m(x)$ is a measurable function of the market parameters as claimed.
\qed

\subsection{Proof of Corollary \ref{cor:mForwardConstruction}}
We first prove the equivalence between $\Phi_i^{(\alpha_1,\dots,\alpha_i)}$ and $\Phi_i^{A_i}$ if in the homogeneous market. Since $(a_j,b_j,c_j)=(a_1,b_1,c_1)$ for all $j=1,\dots,i$, 
\begin{align}
\begin{split}
\Phi_i^{(\alpha_1,\dots,\alpha_i)}(y) & =\frac{(1+b)^{i}}{b^{\sum_{j=1}^i\alpha_j}}\bigg(\sum_{n_1=0,...,n_i=0}^{\infty}(-1)^{p_{(n_1,\dots,n_i)}}Q_{(\alpha_1,\dots,\alpha_i);(n_1,\dots,n_i)}I_0\left(R_{(\alpha_1,\dots,\alpha_i);(n_1,\dots,n_i)}ay\right)\\ & \hspace{2.2cm} -b\sum_{n_1=0,...,n_i=0}^{\infty}(-1)^{p_{(n_1,\dots,n_i)}}Q_{(\alpha_1,\dots,\alpha_i);(n_1,\dots,n_i)}I_0\left(R_{(\alpha_1,\dots,\alpha_i);(n_1,\dots,n_i)}y\right) \bigg),
\end{split}
\end{align}
where
  $Q_{(\alpha_1,\dots,\alpha_i);(n_1,\dots,n_i)}=b^{\sum_{k=1}^in_k(1-2\alpha_k)}$, $R_{(\alpha_1,\dots,\alpha_i);(n_1,\dots,n_i)}=a^{\sum_{s=1}^in_s(2\alpha_s-1)+(\alpha_s-1)}c^{i+1}$, it is enough to simply use $A_i=\sum_{k=1}^i\alpha_k$ to track the process of identifying cases between $(C1)$ and $(C2)$, and we have the simplified form $Q_{(\alpha_1,\dots,\alpha_i);(n_1,\dots,n_i)}=b^{ -\sum\nolimits_{k=1}^{A_i} n_k + \sum\nolimits_{k=A_i+1}^{i} n_k}$, $R_{(\alpha_1,\dots,\alpha_i);(n_1,\dots,n_i)}=a^{\sum\nolimits_{k=1}^{A_i} n_k - \sum\nolimits_{k=A_i+1}^{i} (n_k+1)}c^{i+1}$. Hence, the form of $\Phi_i^{(\alpha_1,\dots,\alpha_i)}$ is equivalent to $\Phi_i^{A_i}$ in the case of homogeneous parameters. Similar arguments hold for $\Psi_i^{(\alpha_1,\dots,\alpha_i)}(y)$ and $\Psi_i^{A_i}$. Furthermore, the forward inverse marginal $I_m$ given in (\ref{eq:ForwardInverseMarginal_Heterogeneous}) can reduce to (\ref{eq:ForwardInverseMarginal_Homogeneous}) with the newly defined $q_{A_m;(n_1,\dots,n_m)}$ and $r_{A_m;(n_1,\dots,n_m)}$.
\qed

\subsection{Proof of Proposition \ref{prop:Connection}}
By virtue of Theorem \ref{thm:mForwardConstructionHeterogeneous}, $U_m$ exists and is unique. 
The single-period forward process $\tilde U_i$ exists for $i =0, \dots, m$ because from the proof of Theorem \ref{thm:mForwardConstructionHeterogeneous} we know that the associated equations of inverse marginal functions $(I_0,\tilde I_m)$, $\tilde I_i(a_iy)+b_i\tilde I_i(y)=(1+b_i)\tilde I_{i-1}(c_iy)$, $i=1,2,\dots,m$ can be solved sequentially for the single-period inverse problems, and $I_m=\tilde I_m$.
On the other hand, if $\tilde U_i$,  $i =0, \dots, m$, is a single-period forward process with $\tilde U_0 = U_0$, with associated optimal wealth process $\tilde X^*$, then 
{\small \begin{align*}
    U_0 (x) = \mathbb{E} \left[ \tilde U_1 ( \tilde X^*_1 ) \right] = \mathbb{E} \left[ \mathbb{E} \left[ \tilde U_2 ( \tilde X^*_2 ) \vert \mathcal{F}_1 \right] \right] = \dots = \mathbb{E} \left[  \tilde U_m ( \tilde X^*_m ) \right]
\end{align*}}
and, with a similar argument, $U_0 (x) \geq \mathbb{E} \left[  \tilde U_m ( \tilde X_m ) \right]$ for any $\tilde X \in \mathcal{X}(x)$. 
Therefore, $(U_0,\tilde U_m)$ is an $m$-forward pair and, by the uniqueness established in Theorem \ref{thm:mForwardConstructionHeterogeneous}, $U_m = \tilde U_m$. Furthermore, Theorem 2.0 in \cite{kramkov1999asymptotic} yields that 
\begin{align*}
\tilde X_m^{*}(x)&=\tilde I_m \left(\rho_m\tilde U_{m-1}^{'} \left(\tilde X_{m-1}^{*}(x) \right) \right)\\
&=\tilde I_m \left(\rho_m\tilde U_{m-1}^{'} \left( \tilde I_{m-1} \left(\rho_{m-1}\tilde U_{m-2}^{'} \left(\tilde X_{m-2}^{*}(x) \right) \right) \right) \right)\\
&=\tilde I_m \left(\rho_m\rho_{m-1}\tilde U_{m-2}^{'} \left(\tilde X_{m-2}^{*}(x) \right) \right)\\
&=\tilde I_m \left(\rho_m\rho_{m-1} \times \dots \times \rho_1U_0^{'}(x) \right)\\
&=I_m \left( \frac{d\Q}{d\P}U_0^{'}(x) \right),
\end{align*}
where for $i=1,2,...,m$, $\rho_i=\frac{q_i}{p_i}\mathbbm{1}_{\left\{R_i=u\right\}}+\frac{1-q_i}{1-p_i}\mathbbm{1}_{\left\{R_i=d\right\}}$. 
\qed

\subsection{Proof of Proposition \ref{prop:BoundsPerformanceSchedule}}
We only discuss the case  $\theta>1$. 
Similar arguments apply when $0<\theta<1$.

Let  $(D_i)_{i=0,\dots, T}$ and $m \in \mathbb{N}$ be fixed.
We first compute the expected performance at times before the first interaction with the robo-advisor. 
By Theorem \ref{thm:mForwardConstructionHeterogeneous} and Corollary \ref{cor:mForwardConstruction}, we have
{\small{\begin{align*}
    \E_1 & \left[ U^{(1)}_i ({X}^{(m)}_i)\big\vert \mathcal{F}_{i-1} \right]\\
    & = D_i p U^{(1)}_i \left(I_i^{(m)} \left( \frac{q}{p}U_{i-1}^{(m)'}({X}^{(m)}_{i-1}) \right) \right)+ (1-D_ip) U^{(1)}_i \left(I_i^{(m)} \left(\frac{1-q}{1-p}U_{i-1}^{(m)'}({X}^{(m)}_{i-1}) \right) \right)\\
    & = D_ip\prod \limits_{j=1}^i\delta_{D_j}^{\frac{1}{\theta}}U_0 \left(I_i^{(m)} \left( \frac{q}{p}U_{i-1}^{(m)'}({X}^{(m)}_{i-1}) \right) \right) \\
    & \qquad +(1-D_ip)\prod \limits_{j=1}^i\delta_{D_j}^{\frac{1}{\theta}}U_0 \left(I_i^{(m)} \left(\frac{1-q}{1-p}U_{i-1}^{(m)'}({X}^{(m)}_{i-1}) \right) \right)\\
    & = D_ip\prod \limits_{j=1}^i\delta_{D_j}^{\frac{1}{\theta}} \left(1-\frac{1}{\theta} \right)^{-1}\delta^{1-\frac{1}{\theta}}q^{1-{\theta}}p^{{\theta}-1} \left({X}^{(m)}_{i-1} \right)^{1-\frac{1}{\theta}}\\ 
    & \qquad + (1-D_ip)\prod\limits_{j=1}^i\delta_{D_j}^{\frac{1}{\theta}} \left(1-\frac{1}{\theta} \right)^{-1}\delta^{1-\frac{1}{\theta}}(1-q)^{1-{\theta}}(1-p)^{{\theta}-1} \left({X}^{(m)}_{i-1} \right)^{1-\frac{1}{\theta}}\\
    & = \delta_{D_i}^{\frac{1}{\theta}}\delta^{1-\frac{1}{\theta}}U_{i-1}^{(1)} ({X}^{(m)}_{i-1}) \left( D_ip^{
    \theta}q^{1-{\theta}}+ \left(\frac{1-D_ip}{1-p} \right) (1-p)^{\theta}(1-q)^{1-\theta} \right),
\end{align*}}}
for $i = 1, \dots, m$, where $\delta_{D_j} = \frac{1+b}{c_j^\theta(a_j^{-\theta}+b)} = \frac{1}{(D_jp)^{\theta}q^{1-\theta} + (1-D_jp)^{\theta}(1-q)^{1-\theta}}$, $j=1,2,\dots,i$, and $\delta=\delta_{D_1}$. Indeed, $U_{i-1}^{(1)} ({X}^{(m)}_{i-1})=\prod \limits_{j=1}^{i-1}\delta_{D_j}^{1-\frac{1}{\theta}}U_0(X_{i-1}^{(m)})=(1-\frac{1}{\theta})^{-1}\prod \limits_{j=1}^{i-1}\delta_{D_j}^{1-\frac{1}{\theta}}(X_{i-1}^{(m)})^{1-\frac{1}{\theta}}$. 

Let $C_1=p^{\theta}q^{1-\theta}$, $C_2=(1-p)^{\theta}(1-q)^{1-\theta}$, and consider a new variable $t_i=(\frac{D_i(1-p)}{1-D_ip})^{\theta}$. 
Clearly, $t_i$ is strictly positive and increasing in $D_i$. 
Then $\E_1 [ U^{(1)}_i ({X}^{(m)}_i)\big\vert \mathcal{F}_{i-1}]$ can be represented by
{{\begin{align*}
    \E_1 [ U^{(1)}_i ({X}^{(m)}_i)\big\vert \mathcal{F}_{i-1}] = \left( \frac{C_1}{(C_1+t_i^{-1}C_2)^{\frac{1}{\theta}}}+\frac{C_2}{(t_iC_1+C_2)^{\frac{1}{\theta}}}\right) U^{(1)}_{i-1} ({X}^{(m)}_{i-1})\delta^{1-\frac{1}{\theta}}.
\end{align*}
}}
Let $f(t_i)=\frac{C_1}{(C_1+t_i^{-1}C_2)^{\frac{1}{\theta}}}+\frac{C_2}{(t_iC_1+C_2)^{\frac{1}{\theta}}}$. After taking derivative we have
\begin{align*}
    f^{'}(t_i)=&\frac{1}{\theta}C_1C_2t_i^{-2}(C_1+t_i^{-1}C_2)^{-\frac{1}{\theta}-1}-\frac{1}{\theta}C_1C_2(t_iC_1+C_2)^{-\frac{1}{\theta}-1}\\=&\frac{1}{\theta}C_1C_2(t_iC_1+C_2)^{-      \frac{1}{\theta}-1}(t_i^{\frac{1}{\theta}-1}-1).
\end{align*}
When $\theta>1$, $U_0(x)>0$, $\frac{1}{\theta}-1<0$, $f(t_i)$ is increasing first and attains its maximum at $t_i=1$, which corresponds to $D_i=1$, and then begins to decrease. Since $f(1)=\frac{C_1+C_2}{(C_1+C_2)^{\frac{1}{\theta}}}=\delta^{\frac{1}{\theta}-1}$, we have $\E_1 [ U^{(1)}_i ({X}^{(m)}_i)\big\vert \mathcal{F}_{i-1}]=U^{(1)}_{i-1} ({X}^{(m)}_{i-1})$ when $D_i=1$. Therefore, let $t_{i,\rm{max}}=(\frac{D_{i,u}(1-p)}{1-D_{i,u}p})^{\theta}$, $t_{i,\rm{min}}=(\frac{D_{i,d}(1-p)}{1-D_{i,d}p})^{\theta}$, $f_{D_{i,u}}=\frac{C_1}{(C_1+t_{i,\rm{max}}^{-1}C_2)^{\frac{1}{\theta}}}+\frac{C_2}{(t_{i,\rm{max}}C_1+C_2)^{\frac{1}{\theta}}}$ and $f_{D_{i,d}}=\frac{C_1}{(C_1+t_{i,\rm{min}}^{-1}C_2)^{\frac{1}{\theta}}}+\frac{C_2}{(t_{i,\rm{min}}C_1+C_2)^{\frac{1}{\theta}}}$. The value range of $f(t_i)$ is $[{\rm{min}}\{f_{D_{i,u}},f_{D_{i,d}}\},\delta^{\frac{1}{\theta}-1}]$, and $\E_1 [ U^{(1)}_i ({X}^{(m)}_i)\big\vert \mathcal{F}_{i-1}]$ is thus bounded between $[\delta^{1-\frac{1}{\theta}}{\rm{min}}\{f_{D_{i,u}},f_{D_{i,d}}\} U^{(1)}_{i-1} ({X}^{(m)}_{i-1}), U^{(1)}_{i-1} ({X}^{(m)}_{i-1})]$ for any possible value of $D_i$ in the interval $[D_{i,d},D_{i,u}]$.

Let $f_j={\rm{min}}\{f_{D_{j,u}},f_{D_{j,d}}\}$ for $j=2,3,...,m$. 
According to the above we have  
\begin{align*}
    \E \left[ U^{(1)}_m (X^{(m)}_m) \right] \leq \E \left[ U^{(1)}_{m-1} (X^{(m)}_{m-1}) \right] \leq \dots \leq U_0 (x)
\end{align*}
and 
\begin{align*}
    \E \left[ U^{(1)}_m (X^{(m)}_m) \right] \geq \E \left[ \delta^{1-\frac{1}{\theta}} f_m U^{(1)}_{m-1}  ({X}^{(m)}_{m-1}) \right] \geq \dots \geq \prod \limits_{j=1}^m f_j \delta^{m(1-\frac{1}{\theta})}U_0(x) .
\end{align*}


When $T > m$ then, according to our assumptions, the agent interacts at time $m$ with the robo-advisor to update $p_m$ back to the original $p$ and her wealth is reduced from $X^{m}_m$ to $\alpha X^{m}_m$. 
Since $U^{(1)}_i ( \alpha x ) = \alpha^{1- \frac{1}{\theta}}U^{(1)}_i ( x )$, which one can show easily by Theorem \ref{thm:mForwardConstructionHeterogeneous}, we can repeat the above steps and obtain that 
\begin{align*}
    \E \left[ U^{(1)}_{2m} (X^{(m)}_{2m}) \right] &\in \left[ \alpha^{1- \frac{1}{\theta}} \prod \limits_{j=1}^{2m} f_j \delta^{2m(1-\frac{1}{\theta})} U_0(x) ,  \alpha^{1- \frac{1}{\theta}}  U_0 (x) \right]\\&=\left[ \alpha^{1- \frac{1}{\theta}} (\prod \limits_{j=1}^{m} f_j)^{2} \delta^{2m(1-\frac{1}{\theta})} U_0(x) ,  \alpha^{1- \frac{1}{\theta}}  U_0 (x) \right].
\end{align*}
The last equality holds because the choice of intervals are periodic. Repeating the above argument then immediately proves the claim. 

To show that $f_i$ is non-increasing in $D_{i,u}$ and non-decreasing in $D_{i,d}$, we notice that $D_{i,u}\geq1$ and thus $t_{i,{\rm{max}}}\geq1$ and that $t_{i,{\rm{max}}}$ is increasing in $D_{i,u}$. 
Therefore, due to the fact that $f(t)$ is decreasing when $t \geq 1$, $f_{D_{i,u}}=f(t_{i,{\rm{max}}})$ is decreasing in $D_{i,u}$. 
One can show analogously that $f_{D_{i,d}}=f(t_{i,{\rm{min}}})$ is increasing in $D_{i,d}$.
Because $f_{D_{i,u}}$ does not depend on $D_{i,d}$ and $f_{D_{i,d}}$ does not depend on $D_{i,u}$, we conclude that $f_i={\rm{min}}\{f_{D_{i,u}},f_{D_{i,d}}\}$ is non-decreasing in $D_{i,u}$ and non-increasing in $D_{i,d}$. 
\qed

\subsection{Proof of Proposition \ref{prop:IndependentT}}
{
We only show the proof for $\theta>1$, similar arguments hold for $0<\theta<1$, but note that then $U_0(x)$ is negative.

Let terminal time $T\in \mathbb{N}$ be given, its minimal expected performance for any interaction schedule $m$ is given by 
\begin{align*}
    \alpha^{(\frac{T}{m}-1)(1-\frac{1}{\theta})}(\prod\limits_{j=1}^{m}f_j)^{\frac{T}{m}}\delta^{T(1-\frac{1}{\theta})}U_0(x)=\frac{\left(\alpha^{1-\frac{1}{\theta}}\prod\limits_{j=1}^{m}(f_j\delta^{1-\frac{1}{\theta}})\right)^{\frac{T}{m}}}{\alpha^{1-\frac{1}{\theta}}}U_0(x)
\end{align*}

Apparently, maximising the minimal expected performance or $\left(\alpha^{1-\frac{1}{\theta}}\prod\limits_{j=1}^{m}(f_j\delta^{1-\frac{1}{\theta}})\right)^{\frac{T}{m}}$ over the divisor $m$ of $T$ is equivalent to maximising $\left(\alpha^{1-\frac{1}{\theta}}\prod\limits_{j=1}^{m}(f_j\delta^{1-\frac{1}{\theta}})\right)^{\frac{1}{m}}$ which is positive for any interaction schedule $m$, and we denote this optimal schedule by $m^{*}$, by considering a large enough $T$, it can be ensured that $\left(\alpha^{1-\frac{1}{\theta}}\prod\limits_{j=1}^{m}(f_j\delta^{1-\frac{1}{\theta}})\right)^{\frac{1}{m}}$ will not continue increasing after $T$ if $m^{*}=T$. The assertion for the independence of $T$ is thus shown, i.e., any setup with horizon $T$ which is a multiple of $m^{*}$ must share the same optimal interaction schedule.
}
\qed

\subsection{Proof of Proposition \ref{prop:BoundDeviationSchedule}}

Let  $(D_i)_{i\in \mathbb{N}}, (E_i)_{i\in \mathbb{N}}$ and $i \in \mathbb{N}$ be fixed.
We first compute
the 1-forward optimal strategy for period $[i-1, i)$  in the ideal world scenario,
\begin{align*}
    \pi_i^{I}(x)=\frac{X_i^{*,u}(x)-X_i^{*,d}(x)}{u-d}=\frac{x}{u-d}\frac{p_i^{\theta_i} q^{-\theta_i}-(1-p_i)^{\theta_i}(1-q)^{-\theta_i}}{p_i^{\theta_i} q^{1-\theta_i}+(1-p_i)^{\theta_i}(1-q)^{1-\theta_i}}, \quad i = 1,\dots, m.
\end{align*}
Next, we give the optimal strategy $\pi^{mF}$ derived in the $m$-forward framework with homogeneous parameters fixed at the beginning of each evaluation period. Since Proposition \ref{prop:Connection} states that the $1$-forward utility is the same as the $m$-forward utility at the same time $m$ when parameters are deterministic and homogeneous, we immediately obtain
\begin{align*}
\begin{split}
    \pi^{mF}_i(x)=\frac{x}{u-d}\frac{q^{-\theta}p^{\theta}-(1-q)^{-\theta}(1-p)^{\theta}}{q^{1-\theta}p^{\theta}+(1-q)^{1-\theta}(1-p)^{\theta}}  \quad i = 1,\dots, m.
\end{split}
\end{align*}
Thus the difference is 
\begin{align*}
    \left\vert\pi^I_i (x) - \pi^{m F}_i (x)\right\vert=\frac{x}{u-d}\left\vert\frac{p_i^{\theta_i} q^{-\theta_i}-(1-p_i)^{\theta_i}(1-q)^{-\theta_i}}{p_i^{\theta_i} q^{1-\theta_i}+(1-p_i)^{\theta_i}(1-q)^{1-\theta_i}}-\frac{q^{-\theta}p^{\theta}-(1-q)^{-\theta}(1-p)^{\theta}}{q^{1-\theta}p^{\theta}+(1-q)^{1-\theta}(1-p)^{\theta}}\right\vert.
\end{align*}
We claim that $\boldsymbol G(\tilde{p},\tilde{\theta})$ is monotone in $\tilde{p}$ and $\tilde{\theta}$. 
Indeed, its first order partial derivative with respect to $\tilde{p}$ is $\boldsymbol G^{'}_{\tilde{p}}(\tilde{p},\tilde{\theta})=\frac{\tilde{\theta} \tilde{p}^{\tilde{\theta}-1}(1-\tilde{p})^{\tilde{\theta}-1}q^{-\tilde{\theta}}(1-q)^{-\tilde{\theta}}}{\left(q^{1-\tilde{\theta}}\tilde{p}^{\tilde{\theta}}+(1-q)^{1-\tilde{\theta}}(1-\tilde{p})^{\tilde{\theta}}\right)^2}>0$. Furthermore, its partial derivative w.r.t. $\tilde{\theta}$ is $\boldsymbol G^{'}_{\tilde{\theta}}(\tilde{p},\tilde{\theta})=\frac{\left(\frac{\tilde{p}}{q}\right)^{\tilde{\theta}}\left(\frac{1-\tilde{p}}{1-q}\right)^{\tilde{\theta}}\left(\ln\frac{\tilde{p}}{q}-\ln\frac{1-\tilde{p}}{1-q}\right)}{\left(q\left(\frac{\tilde{p}}{q}\right)^{\tilde{\theta}}+(1-q)\left(\frac{1-\tilde{p}}{1-q}\right)^{\tilde{\theta}}\right)^2}$, which is strictly positive when $\tilde{p}>q$ and strictly negative when $\tilde{p}<q$. Overall, the largest possible absolute deviation is attained by one of the following four cases: $(1) \enskip D_i=D_{i,u}, E_i=E_{i,u}$ if $D_{i,d}p<q$, $(2) \enskip D_i=D_{i,u}, E_i=E_{i,d}$ if $D_{i,u}p>q$, $(3) \enskip D_i=D_{i,d}, E_i=E_{i,d}$ if $D_{i,d}p<q$ or $(4) \enskip D_i=D_{i,d}, E_i=E_{i,u}$ if $D_{i,d}p<q$. The essential supremum and the monotonicity in $i$ can be naturally derived then.

\end{appendices}
\bibliography{BibFile}

\begin{thebibliography}{49}
\providecommand{\natexlab}[1]{#1}
\providecommand{\url}[1]{\texttt{#1}}
\expandafter\ifx\csname urlstyle\endcsname\relax
  \providecommand{\doi}[1]{doi: #1}\else
  \providecommand{\doi}{doi: \begingroup \urlstyle{rm}\Url}\fi

\bibitem[Aldridge(2013)]{aldridge2013high}
Irene Aldridge.
\newblock \emph{High-frequency trading: a practical guide to algorithmic
  strategies and trading systems}, volume 604.
\newblock John Wiley \& Sons, 2013.

\bibitem[Angoshtari(2022)]{angoshtari2022predictable}
Bahman Angoshtari.
\newblock Predictable forward performance processes in complete markets.
\newblock \emph{Probability Uncertainty and Quantitative Risk, accepted}, 2022.

\bibitem[Angoshtari et~al.(2020)Angoshtari, Zariphopoulou, and
  Zhou]{angoshtari2020predictable}
Bahman Angoshtari, Thaleia Zariphopoulou, and Xun~Yu Zhou.
\newblock Predictable forward performance processes: The binomial case.
\newblock \emph{SIAM Journal on Control and Optimization}, 58\penalty0
  (1):\penalty0 327--347, 2020.

\bibitem[Avanesyan et~al.(2020)Avanesyan, Shkolnikov, and
  Sircar]{Avanesyan2020}
Levon Avanesyan, Mykhaylo Shkolnikov, and Ronnie Sircar.
\newblock Construction of a class of forward performance processes in
  stochastic factor models, and an extension of widder theorem.
\newblock \emph{Finance and Stochastics}, 24\penalty0 (4):\penalty0 981–1011,
  2020.

\bibitem[Berrier et~al.(2009)Berrier, Rogers, and
  Tehranchi]{rogers2009characterization}
Francois Berrier, Leonard~CG Rogers, and Michael Tehranchi.
\newblock A characterization of forward utility functions.
\newblock \emph{Statistical Laboratory, University of Cambridge, Cambridge,
  UK}, 2009.

\bibitem[Bo et~al.(2022)Bo, Capponi, and Zhou]{Bo2022}
Lijun Bo, Agostino Capponi, and Chao Zhou.
\newblock Power forward performance in semimartingale markets with stochastic
  integrated factors.
\newblock \emph{Mathematics of Operations Research, accepted}, 2022.

\bibitem[Brogaard et~al.(2014)Brogaard, Hendershott, and
  Riordan]{brogaard2014high}
Jonathan Brogaard, Terrence Hendershott, and Ryan Riordan.
\newblock High-frequency trading and price discovery.
\newblock \emph{The Review of Financial Studies}, 27\penalty0 (8):\penalty0
  2267--2306, 2014.

\bibitem[Capponi et~al.(2022)Capponi, Olafsson, and
  Zariphopoulou]{capponi2022personalized}
Agostino Capponi, Sveinn Olafsson, and Thaleia Zariphopoulou.
\newblock Personalized robo-advising: Enhancing investment through client
  interaction.
\newblock \emph{Management Science}, 68\penalty0 (4):\penalty0 2485--2512,
  2022.

\bibitem[Cartea et~al.(2015)Cartea, Jaimungal, and
  Penalva]{cartea2015algorithmic}
{\'A}lvaro Cartea, Sebastian Jaimungal, and Jos{\'e} Penalva.
\newblock \emph{Algorithmic and high-frequency trading}.
\newblock Cambridge University Press, 2015.

\bibitem[Chaboud et~al.(2014)Chaboud, Chiquoine, Hjalmarsson, and
  Vega]{chaboud2014rise}
Alain~P Chaboud, Benjamin Chiquoine, Erik Hjalmarsson, and Clara Vega.
\newblock Rise of the machines: Algorithmic trading in the foreign exchange
  market.
\newblock \emph{The Journal of Finance}, 69\penalty0 (5):\penalty0 2045--2084,
  2014.

\bibitem[Chong(2019)]{chong2019pricing}
Wing~Fung Chong.
\newblock Pricing and hedging equity-linked life insurance contracts beyond the
  classical paradigm: The principle of equivalent forward preferences.
\newblock \emph{Insurance: Mathematics and Economics}, 88:\penalty0 93--107,
  2019.

\bibitem[Choulli et~al.(2007)Choulli, Stricker, and Li]{choulli2007minimal}
Tahir Choulli, Christophe Stricker, and Jia Li.
\newblock Minimal hellinger martingale measures of order q.
\newblock \emph{Finance and Stochastics}, 11\penalty0 (3):\penalty0 399--427,
  2007.

\bibitem[Cox et~al.(1979)Cox, Ross, and Rubinstein]{cox1979option}
John~C Cox, Stephen~A Ross, and Mark Rubinstein.
\newblock Option pricing: A simplified approach.
\newblock \emph{Journal of Financial Economics}, 7\penalty0 (3):\penalty0
  229--263, 1979.

\bibitem[Cui et~al.(2022)Cui, Li, Qiao, and Strub]{cui2022risk}
Xiang-Yu Cui, Duan Li, Xiao Qiao, and Moris~S Strub.
\newblock Risk and potential: An asset allocation framework with applications
  to robo-advising.
\newblock \emph{Journal of the Operations Research Society of China},
  10\penalty0 (3):\penalty0 529--558, 2022.

\bibitem[D'Acunto and Rossi(2021)]{DAcuntoRossi21:Inbook}
Francesco D'Acunto and Alberto~G. Rossi.
\newblock Robo-advising.
\newblock In Raghavendra Rau, Robert Wardrop, and Luigi Zingales, editors,
  \emph{The Palgrave Handbook of Technological Finance}, page 725–749.
  Palgrave Macmillan Cham, 2021.

\bibitem[Dai et~al.(2021{\natexlab{a}})Dai, Jin, Kou, and Xu]{dai2021dynamic}
Min Dai, Hanqing Jin, Steven Kou, and Yuhong Xu.
\newblock A dynamic mean-variance analysis for log returns.
\newblock \emph{Management Science}, 67\penalty0 (2):\penalty0 1093--1108,
  2021{\natexlab{a}}.

\bibitem[Dai et~al.(2021{\natexlab{b}})Dai, Jin, Kou, and Xu]{dai2021robo}
Min Dai, Hanqing Jin, Steven Kou, and Yuhong Xu.
\newblock Robo-advising: A dynamic mean-variance approach.
\newblock \emph{Digital Finance}, pages 1--17, 2021{\natexlab{b}}.

\bibitem[El~Karoui and Mrad(2013)]{mohamed2013exact}
Nicole El~Karoui and Mohamed Mrad.
\newblock An exact connection between two solvable sdes and a nonlinear utility
  stochastic pde.
\newblock \emph{SIAM Journal on Financial Mathematics}, 4\penalty0
  (1):\penalty0 697, 2013.

\bibitem[El~Karoui and Mrad(2021)]{el2021recover}
Nicole El~Karoui and Mohamed Mrad.
\newblock Recover dynamic utility from observable process: Application to the
  economic equilibrium.
\newblock \emph{SIAM Journal on Financial Mathematics}, 12\penalty0
  (1):\penalty0 189--225, 2021.

\bibitem[El~Karoui et~al.(2014)El~Karoui, Hillairet, and Mrad]{el2014}
Nicole El~Karoui, Caroline Hillairet, and Mohamed Mrad.
\newblock Affine long term yield curves: an application of the {R}amsey rule
  with progressive utility.
\newblock \emph{International Journal of Financial Engineering}, 1\penalty0
  (1):\penalty0 1450003, 24, 2014.

\bibitem[El~Karoui et~al.(2018)El~Karoui, Hillairet, and
  Mrad]{el2018consistent}
Nicole El~Karoui, Caroline Hillairet, and Mohamed Mrad.
\newblock Consistent utility of investment and consumption: a forward/backward
  spde viewpoint.
\newblock \emph{Stochastics}, 90\penalty0 (6):\penalty0 927--954, 2018.

\bibitem[El~Karoui et~al.(2022)El~Karoui, Hillairet, and Mrad]{el2021}
Nicole El~Karoui, Caroline Hillairet, and Mohamed Mrad.
\newblock Ramsey rule with forward/backward utility for long-term yield curves
  modeling.
\newblock \emph{Decisions in Economics and Finance}, 45\penalty0 (1):\penalty0
  375--414, 2022.

\bibitem[Fabozzi et~al.(2010)Fabozzi, Focardi, and
  Kolm]{fabozzi2010quantitative}
Frank~J Fabozzi, Sergio~M Focardi, and Petter~N Kolm.
\newblock \emph{Quantitative equity investing: Techniques and strategies}.
\newblock John Wiley \& Sons, 2010.

\bibitem[Geng and Zariphopoulou(2017)]{geng2017temporal}
Tianran Geng and Thaleia Zariphopoulou.
\newblock Temporal and spatial turnpike-type results under forward
  time-monotone performance criteria.
\newblock \emph{arXiv preprint arXiv:1702.05649}, 2017.

\bibitem[Grossman(1988)]{grossman1988program}
Sanford~J Grossman.
\newblock Program trading and market volatility: A report on interday
  relationships.
\newblock \emph{Financial Analysts Journal}, 44\penalty0 (4):\penalty0 18--28,
  1988.

\bibitem[He et~al.(2021)He, Strub, and Zariphopoulou]{he2021forward}
Xue~Dong He, Moris~S Strub, and Thaleia Zariphopoulou.
\newblock Forward rank-dependent performance criteria: Time-consistent
  investment under probability distortion.
\newblock \emph{Mathematical Finance}, 31\penalty0 (2):\penalty0 683--721,
  2021.

\bibitem[Hendershott et~al.(2011)Hendershott, Jones, and
  Menkveld]{hendershott2011does}
Terrence Hendershott, Charles~M Jones, and Albert~J Menkveld.
\newblock Does algorithmic trading improve liquidity?
\newblock \emph{The Journal of Finance}, 66\penalty0 (1):\penalty0 1--33, 2011.

\bibitem[Henderson and Hobson(2007)]{henderson2007horizon}
Vicky Henderson and David Hobson.
\newblock Horizon-unbiased utility functions.
\newblock \emph{Stochastic Processes and their Applications}, 117\penalty0
  (11):\penalty0 1621--1641, 2007.

\bibitem[Hu et~al.(2020)Hu, Liang, and Tang]{hu2020systems}
Ying Hu, Gechun Liang, and Shanjian Tang.
\newblock Systems of ergodic {BSDE}s arising in regime switching forward
  performance processes.
\newblock \emph{SIAM Journal on Control and Optimization}, 58\penalty0
  (4):\penalty0 2503--2534, 2020.

\bibitem[Jackwerth(1999)]{Jackwerth99:JoD}
Jens~Carsten Jackwerth.
\newblock Option-implied risk-neutral distributions and implied binomial trees.
\newblock \emph{The Journal of Derivatives}, 7\penalty0 (2):\penalty0 66--82,
  1999.

\bibitem[Kallblad(2020)]{k2020black}
Sigrid Kallblad.
\newblock Black's inverse investment problem and forward criteria with
  consumption.
\newblock \emph{SIAM Journal on Financial Mathematics}, 11\penalty0
  (2):\penalty0 494--525, 2020.

\bibitem[K{\"a}llblad et~al.(2018)K{\"a}llblad, Ob{\l}{\'o}j, and
  Zariphopoulou]{kallblad2018dynamically}
Sigrid K{\"a}llblad, Jan Ob{\l}{\'o}j, and Thaleia Zariphopoulou.
\newblock Dynamically consistent investment under model uncertainty: the robust
  forward criteria.
\newblock \emph{Finance and Stochastics}, 22\penalty0 (4):\penalty0 879--918,
  2018.

\bibitem[Kramkov and Schachermayer(1999)]{kramkov1999asymptotic}
Dimitri Kramkov and Walter Schachermayer.
\newblock The asymptotic elasticity of utility functions and optimal investment
  in incomplete markets.
\newblock \emph{Annals of Applied Probability}, pages 904--950, 1999.

\bibitem[Kress et~al.(1989)Kress, Maz'ya, and Kozlov]{kress1989linear}
Rainer Kress, V~Maz'ya, and V~Kozlov.
\newblock \emph{Linear integral equations}, volume~82.
\newblock Springer, 1989.

\bibitem[Kuczma et~al.(1990)Kuczma, Choczewski, and
  Ger]{kuczma_choczewski_ger_1990}
Marek Kuczma, Bogdan Choczewski, and Roman Ger.
\newblock \emph{Iterative Functional Equations}.
\newblock Encyclopedia of Mathematics and its Applications. Cambridge
  University Press, 1990.

\bibitem[Liang and Zariphopoulou(2017)]{liang2017representation}
Gechun Liang and Thaleia Zariphopoulou.
\newblock Representation of homothetic forward performance processes in
  stochastic factor models via ergodic and infinite horizon {BSDE}.
\newblock \emph{SIAM Journal on Financial Mathematics}, 8\penalty0
  (1):\penalty0 344--372, 2017.

\bibitem[Menkveld(2013)]{menkveld2013high}
Albert~J Menkveld.
\newblock High frequency trading and the new market makers.
\newblock \emph{Journal of Financial Markets}, 16\penalty0 (4):\penalty0
  712--740, 2013.

\bibitem[Musiela and Zariphopoulou(2006)]{musiela2006investments}
Marek Musiela and Thaleia Zariphopoulou.
\newblock Investments and forward utilities.
\newblock \emph{Preprint}, 2006.

\bibitem[Musiela and Zariphopoulou(2008)]{musiela2008optimal}
Marek Musiela and Thaleia Zariphopoulou.
\newblock Optimal asset allocation under forward exponential performance
  criteria.
\newblock \emph{Markov processes and related topics: a Festschrift for Thomas
  G. Kurtz}, 4:\penalty0 285--300, 2008.

\bibitem[Musiela and Zariphopoulou(2009)]{musiela2009portfolio}
Marek Musiela and Thaleia Zariphopoulou.
\newblock Portfolio choice under dynamic investment performance criteria.
\newblock \emph{Quantitative Finance}, 9\penalty0 (2):\penalty0 161--170, 2009.

\bibitem[Musiela and Zariphopoulou(2010)]{musiela2010portfolio}
Marek Musiela and Thaleia Zariphopoulou.
\newblock Portfolio choice under space-time monotone performance criteria.
\newblock \emph{SIAM Journal on Financial Mathematics}, 1\penalty0
  (1):\penalty0 326--365, 2010.

\bibitem[Nadtochiy and Tehranchi(2017)]{nadtochiy2017optimal}
Sergey Nadtochiy and Michael Tehranchi.
\newblock Optimal investment for all time horizons and {M}artin boundary of
  space-time diffusions.
\newblock \emph{Mathematical Finance}, 27\penalty0 (2):\penalty0 438--470,
  2017.

\bibitem[O'hara(2015)]{o2015high}
Maureen O'hara.
\newblock High frequency market microstructure.
\newblock \emph{Journal of Financial Economics}, 116\penalty0 (2):\penalty0
  257--270, 2015.

\bibitem[Polyanin and Manzhirov(2008)]{polyanin2008handbook}
Andrei~D Polyanin and Alexander~V Manzhirov.
\newblock \emph{Handbook of integral equations}.
\newblock Chapman and Hall/CRC, 2008.

\bibitem[Rubinstein(1994)]{Rubinstein94:JF}
Mark Rubinstein.
\newblock Implied binomial trees.
\newblock \emph{The Journal of Finance}, 49\penalty0 (3):\penalty0 771--818,
  1994.

\bibitem[Shkolnikov et~al.(2016)Shkolnikov, Sircar, and
  Zariphopoulou]{shkolnikov2016asymptotic}
Mykhaylo Shkolnikov, Ronnie Sircar, and Thaleia Zariphopoulou.
\newblock Asymptotic analysis of forward performance processes in incomplete
  markets and their ill-posed {HJB} equations.
\newblock \emph{SIAM Journal on Financial Mathematics}, 7\penalty0
  (1):\penalty0 588--618, 2016.

\bibitem[Strub and Zhou(2021)]{strub2021evolution}
Moris~S Strub and Xun~Yu Zhou.
\newblock Evolution of the {A}rrow--{P}ratt measure of risk-tolerance for
  predictable forward utility processes.
\newblock \emph{Finance and Stochastics}, 25\penalty0 (2):\penalty0 331--358,
  2021.

\bibitem[Zemyan(2012)]{zemyan2012classical}
Stephen~M Zemyan.
\newblock \emph{The classical theory of integral equations: a concise
  treatment}.
\newblock Springer Science \& Business Media, 2012.

\bibitem[{\v{Z}}itkovi{\'c}(2009)]{vzitkovic2009dual}
Gordan {\v{Z}}itkovi{\'c}.
\newblock A dual characterization of self-generation and exponential forward
  performances.
\newblock \emph{The Annals of Applied Probability}, 19\penalty0 (6):\penalty0
  2176--2210, 2009.

\end{thebibliography}

\end{document}